\documentclass[11pt, reprint, amsmath, amssymb, aps, prl]{revtex4-2}

\usepackage{mathrsfs}
\usepackage{tikz}
\usepackage{tikz-cd}
\usepackage[usenames, dvipsnames,svgnames]{xcolor}
\usetikzlibrary{decorations.pathmorphing}
\usetikzlibrary{decorations.markings}
\usepackage{graphicx}
\usepackage{float}
\usepackage{scrextend}
\usepackage{manfnt,pifont}
\usepackage{hyperref}
\usepackage{url}
\usepackage{lipsum}
\usepackage[font=small]{subcaption}
\usepackage[font=small]{caption}

\usepackage{dsfont}
\usetikzlibrary{arrows.meta} 
\usepackage{subcaption}
\usepackage{appendix}

\usepackage[normalem]{ulem}

\usepackage{url}
\usepackage[outline]{contour}
\usepackage{textgreek}

\newcommand{\cL}{{\mathcal L}}

\usepackage{tikz}
\usetikzlibrary{shapes}
\tikzset{line/.style={line width=0.25mm},
curve/.style={line,smooth,tension=1},
->-/.style={decoration={
  markings,
  mark=at position #1 with {\arrow[>=stealth]{>}}},postaction={decorate}},
-<-/.style={decoration={
  markings,
  mark=at position #1 with {\arrow[>=stealth]{<}}},postaction={decorate}},
}
\newcommand{\dotsol}[2]{node [circle, opacity=1, fill, inner sep=1pt, label = #1 : {#2}] {}}

\newcommand{\dotsols}[2]{node [circle, fill, inner sep=1pt, label={#1:#2}] {}}

\tikzset{
  sharp arrow/.style={
    -Stealth 
  }
}

\usepackage{CJK}
\usepackage{enumitem}
\setlist[itemize]{leftmargin=*}

\makeatletter
\DeclareFontFamily{OMX}{MnSymbolE}{}
\DeclareSymbolFont{MnLargeSymbols}{OMX}{MnSymbolE}{m}{n}
\SetSymbolFont{MnLargeSymbols}{bold}{OMX}{MnSymbolE}{b}{n}
\DeclareFontShape{OMX}{MnSymbolE}{m}{n}{
    <-6>  MnSymbolE5
   <6-7>  MnSymbolE6
   <7-8>  MnSymbolE7
   <8-9>  MnSymbolE8
   <9-10> MnSymbolE9
  <10-12> MnSymbolE10
  <12->   MnSymbolE12
}{}
\DeclareFontShape{OMX}{MnSymbolE}{b}{n}{
    <-6>  MnSymbolE-Bold5
   <6-7>  MnSymbolE-Bold6
   <7-8>  MnSymbolE-Bold7
   <8-9>  MnSymbolE-Bold8
   <9-10> MnSymbolE-Bold9
  <10-12> MnSymbolE-Bold10
  <12->   MnSymbolE-Bold12
}{}

\let\llangle\@undefined
\let\rrangle\@undefined
\DeclareMathDelimiter{\llangle}{\mathopen}%
                     {MnLargeSymbols}{'164}{MnLargeSymbols}{'164}
\DeclareMathDelimiter{\rrangle}{\mathclose}%
                     {MnLargeSymbols}{'171}{MnLargeSymbols}{'171}

\makeatletter

\def\l@subsection#1#2{}

\def\l@subsubsection#1#2{}

\makeatother

\usepackage{eso-pic}
\newcommand{\reportnum}[2]{
  \AddToShipoutPictureBG*{%
    \AtPageUpperLeft{%
      \hspace{0.75\paperwidth}%
      \raisebox{#1\baselineskip}{%
        \makebox[0pt][l]{\textnormal{#2}}
  }}}%
}

\begin{document}

\reportnum{-3}{USTC-ICTS/PCFT-25-54}

\title{Fermionic Non-invertible Symmetry Behind Supersymmetric \textnormal{\textit{ADE}} Solitons}
\author{Jin Chen$^{a,b}$}
\email{zenofox@gmail.com}
\author{Zhihao Duan$^{c,d}$}
\email{xduanz@gmail.com}
\author{Qiang Jia$^{e}$}
\email{qjia1993@kaist.ac.kr}
\author{Sungjay Lee$^{f}$}
\email{sjlee@kias.re.kr}
\affiliation{$^{a}$Department of Physics, Xiamen University, Xiamen, 361005, China}
\affiliation{$^{b}$Peng Huanwu Center for Fundamental Theory, Hefei, Anhui 230026, China}
\affiliation{$^{c}$Section de Math\'ematiques, Universit\'e de Gen\`eve, 1211 Gen\`eve 4, Switzerland}
\affiliation{$^{d}$Centre for Theoretical Physics, Department of Physics and Astronomy, \\
Queen Mary University of London, London E1 4NS, UK}
\affiliation{$^{e}$Department of Physics, Korea Advanced Institute of Science \& Technology,
Daejeon 34141, Korea}
\affiliation{$^{f}$Korea Institute for Advanced Study, 85 Hoegiro, Dongdaemun-Gu, Seoul 02455, Korea}

\begin{abstract}
The non-perturbative constraints imposed by intrinsic fermionic non-invertible symmetries in 1+1 dimensional gapped systems 
remain largely unexplored. In this letter, we propose the superstrip algebra as a unified 
framework to catalog the categorical symmetry data in a massive fermionic model. 
The algebra and its representations explicitly encode the vacuum structure, soliton degeneracies, and their quantum numbers. As a demonstration, we apply this framework to 
the $\mathcal N=2$ minimal models with their least relevant deformation. We show that this specific deformation alone preserves a non-invertible superfusion category, a fermionic variant of ${\rm SU}(2)_k$ known to underlie the $ADE$ classification of critical theories. 
Its superstrip algebra then accounts for 
the origin of the resulting $ADE$-type soliton spectrum and their fractional fermion number. Although our primary examples are supersymmetric and integrable, our framework itself relies on neither property, providing a new powerful tool for studying a broad class of strongly-coupled fermionic systems. 
\end{abstract}

\maketitle

\section{Introduction}
The dynamics of Quantum Field Theory are fundamentally 
shaped by symmetries and their anomalies. 
They impose powerful constraints on renormalization group (RG) flows, forbidding certain flow directions in the space of couplings. 
Upon adding deformation terms that either break or preserve the very symmetries, the long-distance phase, whether gapped or gapless, 
is then characterized by preserved infrared (IR) symmetries and their associated 't Hooft anomalies. 

Recent generalizations of symmetry, including higher-form and categorical symmetries, have provided powerful constraints and new 
perspectives on strongly-coupled systems 
\cite{Chang:2018iay,Thorngren:2019iar,Komargodski:2020mxz,Thorngren:2021yso,Cordova:2022ieu,Apte:2022xtu,Lin:2023uvm,Kaidi:2023maf,Zhang:2023wlu,Damia:2023ses,Cordova:2023bja,Choi:2023pdp,Antinucci:2023ezl,Bhardwaj:2024kvy,Bhardwaj:2024qiv,Cordova:2024ypu,Copetti:2024rqj,DelZotto:2024arv,Nakayama:2024msv, Cordova:2024goh, Cordova:2024nux,Bhardwaj:2025piv,Bhardwaj:2025jtf,Seiberg:2025bqy,Antinucci:2025fjp}. However, this focus has been overwhelmingly directed at bosonic systems, and studies on fermionic systems remain few \cite{Wan:2016php, Lou:2020gfq, Hu:2021qhm, Aasen:2017ubm, Wang:2023iqt,Ambrosino:2024ggh,Bhardwaj:2024ydc,Balasubramanian:2024nei}. The nature of intrinsically fermionic categorical symmetries \cite{note:intrinsic} (and their anomalies), especially how they constrain non-perturbative physics, remains largely unexplored. 

In the present work, we propose a general framework, called ``superstrip algebra'' to address this gap in 1+1 dimensions (2D). The superstrip algebra is 
built on the spontaneously broken fermionic non-invertible symmetries in a gapped system of fermions. This framework reveals that:

\emph{The non-perturbative spectrum (vacua, particles, and solitons) is only kinematically consistent if it satisfies the novel and stringent selection rules imposed by the underlying superstrip algebra and its quiver representation.}

To demonstrate how useful this framework is, we examine the 2D $\mathcal N=2$ superconformal minimal models with their least relevant deformation.
The minimal conformal models are well-known to be classified by the $ADE$ series, originated from the modular invariant partition functions of the $\hat A_1$ Kac-Moody algebra \cite{Cappelli:1987xt}. This classification finds a modern interpretation in terms of 
generalized symmetry. The distinct $ADE$ theories are related by gauging (non-)invertible symmetries \cite{Fuchs:2002cm, Diatlyk:2023fwf, Tanaka:2025qou}, and their topological defect lines are organized according to the corresponding Dynkin diagrams. This entire structure is not robust and is easily destroyed by generic deformations. Nonetheless, we demonstrate that the least relevant deformation delicately preserves an $\hat A_1$-like superfusion category. Applying the superstrip algebra, we show that this superfusion category ultimately governs the dynamics of the massive minimal models, leaving the characteristic $ADE$-pattern imprinted on their solitonic spectra.


\section{Superfusion category and superstrip algebra}
\label{Sec: catandalg}



\paragraph{Superfusion category}
In 1+1 dimensions, discrete global symmetries are characterized by a set of topological defect lines (TDLs) \cite{Bhardwaj:2017xup,Aasen:2017ubm,Chang:2018iay,Chang:2022hud}. A TDL $\mathcal{L}$ will be drawn as an oriented line. Its orientation reversal is labeled by $\overline{\mathcal{L}}$. For an ordinary finite group symmetry $G$, a TDL, denoted as $\mathcal{L}_g$ is associated with each group element $g\in G$. Then $\overline{\mathcal{L}_g} = \mathcal{L}_{g^{-1}}$, and the action of $\mathcal{L}_g$ on a charged operator $\mathcal{O}$ can be understood as wrapping the line around $\mathcal{O}$ and shrinking it
\begin{equation}
\begin{gathered}
\begin{tikzpicture}[scale=0.4]
\draw [line,ultra thick,dashed,MidnightBlue] (-.5,-0.75) circle (1.7) ;
\draw (-3.1,-0.75) node {$\mathcal{L}_g$};
\draw (0,1.2) node {};
\draw (-0.5,-0.75)\dotsol {right}{$\mathcal{O}$};
\draw [line,ultra thick,DarkBlue,-<-=1.0](-.47, 0.95) -- (-.46, 0.95);
\draw [line,ultra thick,DarkBlue,-<-=1.0](-.48, -2.45)--(-.49, -2.45);
\end{tikzpicture}
\end{gathered}\quad = \quad
\begin{gathered}
\begin{tikzpicture}[scale=1]
\draw (0,-0.3)\dotsol {below}{$ \widehat{\mathcal{L}}_{g}(\mathcal O)$};
\draw (0,.45) node {};
\end{tikzpicture}
\end{gathered}  
\end{equation}
and the group law is encoded in the fusion of two TDLs.
Considering TDLs as symmetry generators generalizes the conventional notion of symmetry. In the example of Ising conformal field theory (CFT), the Kramers-Wannier duality is a symmetry that is not group-like under fusion. To fully capture all possible symmetries, a more general framework is necessary to encode the rich properties of TDLs. Taking into account fermions, we are led to the notion of a superfusion category $\mathscr{C}$. We summarize some key properties and highlight those that contain new features compared to the fusion category:

\begin{itemize}
    \item $\mathbf{Direct\, Sum}$: Given two TDLs $\mathcal{L}_a$ and $\mathcal{L}_b$, we can define the sum $\mathcal{L}_a + \mathcal{L}_b$
        . A \textit{simple} TDL cannot be further decomposed and represents a simple object in $\mathscr{C}$. We assume that all TDLs can be decomposed into a direct sum of simple TDLs, and the set of simple TDLs is finite.

    \item $\mathbf{Junction}$: A set of TDLs $\{\mathcal{L}_a,\mathcal{L}_b,\cdots\}$ can meet at a junction point, and we associate to it a junction vector space $V_{\mathcal{L}_a,\mathcal{L}_b,\cdots}$ which defines morphisms in $\mathscr{C}$. In the case of CFT, $V_{\mathcal{L}_a,\mathcal{L}_b,\cdots}$ is identified with the vacuum subspace (weight (0,0) states) of the defect Hilbert space $\mathcal{H}_{\mathcal{L}_a,\mathcal{L}_b,\cdots}$ under radial quantization.

    \item[\textcolor{red}{\textbullet}] $\mathbf{Grading}$: The junction vector space admits a $\mathbb{Z}_2$ grading, which dictates the bosonic or fermionic nature of the junction. 
    A typical vector space takes the form $\mathbb{C}^{n|m}$, with $n$-dimensional bosonic subspace and $m$-dimensional fermionic subspace. Physically, a TDL $\mathcal{L}$ is simple if the junction vector space $V_{\mathcal{L},\overline{\mathcal{L}}}$ is isomorphic to either $\mathbb{C}^{1|0}$ or $\mathbb{C}^{1|1}$. The former defines an \emph{$m$-type} TDL, while the latter is referred to as \emph{$q$-type} TDL.  
For example, the fermion parity TDL $(-1)^F$ is known to be of $m$-type. A $q$-type TDL can host a 1D Majorana fermion, represented by a red dot, which accounts for the fermionic subspace in $\mathbb{C}^{1|1}$. Two red dots can meet and annihilate along a $q$-type line, and exchanging their positions introduces a minus sign.
\begin{equation}
    \begin{gathered}
        \begin{tikzpicture}[scale=0.8]
    \draw[ultra thick,dashed,MidnightBlue,-<-=.55] (3.25,-1)--(3.25,1);
    \node at (4,0) {=};
    \draw[ultra thick,dashed,MidnightBlue,-<-=.55] (4.75,-1)--(4.75,1);
    \filldraw[red] (4.75,-0.5) circle (2pt);
    \filldraw[red] (4.75,0.5) circle (2pt);
    \node at (4.5,-0.5) {$1$};
    \node at (4.5,0.5) {$2$};
    \node at (5.5,0) {$=$};
    \node at (6,0) {$-$};
    \draw[ultra thick,dashed,MidnightBlue,-<-=.55] (4.75+2,-1)--(4.75+2,1);
    \filldraw[red] (4.75+2,-0.5) circle (2pt);
    \filldraw[red] (4.75+2,0.5) circle (2pt);
    \node at (4.5+2,-0.5) {$2$};
    \node at (4.5+2,0.5) {$1$};
    \end{tikzpicture}
    \end{gathered}\quad \,, \quad
    \begin{gathered}
        \begin{tikzpicture}[scale=0.8]
    \draw[ultra thick,dashed,MidnightBlue,-<-=.55] (3.5,-1)--(3.5,1);
    \draw[ultra thick,dashed,MidnightBlue,-<-=.55] (4.5,-1)--(4.5,1);
    \filldraw[red] (4.5,-0.5) circle (2pt);
    \filldraw[red] (3.5,0.5) circle (2pt);
    \end{tikzpicture}
    \end{gathered} \quad =\quad - \quad
    \begin{gathered}
        \begin{tikzpicture}[scale=0.8]
    \draw[ultra thick,dashed,MidnightBlue,-<-=.55] (3.5,-1)--(3.5,1);
    \draw[ultra thick,dashed,MidnightBlue,-<-=.55] (4.5,-1)--(4.5,1);
    \filldraw[red] (4.5,0.5) circle (2pt);
    \filldraw[red] (3.5,-0.5) circle (2pt);
    \end{tikzpicture}
    \end{gathered}    
\end{equation}

    \item[\textcolor{red}{\textbullet}] $\mathbf{Fusion}$: Two TDLs $\mathcal{L}_a$ and $\mathcal{L}_b$ can fuse to a new TDL, which is then further decomposed into a sum of simple TDLs. This process is dictated by the fusion rule,
\begin{equation}
    \mathcal{L}_a \cdot \mathcal{L}_b = \sum_{c} N_{ab}^{c,\mathbf{b}} \mathcal{L}_c + \sum_{c} N_{ab}^{c,\mathbf{f}} \mathcal{L}_c\,,
\end{equation}
where $N_{ab}^{c,\mathbf{b}(\mathbf{f})}$ are non-negative integers reflecting the dimension of the junction vector space
    \begin{equation}
        V_{\mathcal{L}_a,\mathcal{L}_b,\overline{\mathcal{L}}_c} = \mathbb{C}^{N_{ab}^{c,\mathbf{b}}|N_{ab}^{c,\mathbf{f}}}\,,
    \end{equation}
where we use $\mathbf{b}$ and $\mathbf{f}$ to denote bosonic and fermionic degrees of freedom respectively. Pictorially, we have
\begin{equation}
    \begin{gathered}
\begin{tikzpicture}[scale=0.8]
\draw [line,dashed,ultra thick,MidnightBlue,->-=.55] (0,0)  -- (0,-1) node [below=-3pt] {$\cL_c$};
\draw [line,dashed,ultra thick,BurntOrange,-<-=.55] (0,0) -- (-1,1) node [above=-3pt] {$\cL_a$};
\draw [line,dashed,ultra thick,ForestGreen,-<-=.55] (0,0) --(1,1) node [above=-3pt] {$\cL_b$};
\node at (0.5,0) {{\color{black}{$\alpha^{\mathbf{b}}$}}};
\filldraw[Cerulean] (0,0) circle (3pt);
\end{tikzpicture}
\end{gathered} \quad
    \begin{gathered}
\begin{tikzpicture}[scale=0.8]
\draw [line,dashed,ultra thick,MidnightBlue,->-=.55] (0,0)  -- (0,-1) node [below=-3pt] {$\cL_c$};
\draw [line,dashed,ultra thick,BurntOrange,-<-=.55] (0,0) -- (-1,1) node [above=-3pt] {$\cL_a$};
\draw [line,dashed,ultra thick,ForestGreen,-<-=.55] (0,0) --(1,1) node [above=-3pt] {$\cL_b$};
\node at (0.5,0) {{\color{black}{$\alpha^{\mathbf{f}}$}}};
\filldraw[red] (0,0) circle (3pt);
\end{tikzpicture}
\end{gathered}
\end{equation}    
with $\alpha^{\bf b}=1,\cdots,N_{ab}^{c,\bf b}$ and $\alpha^{\bf f}=1,\cdots,N_{ab}^{c,\bf f}$.

    \item[\textcolor{red}{\textbullet}] $\mathbf{Crossing}$: Given a network of TDLs, we can consistently transform one to another by a sequence of crossing relations or $F$-moves. 
Consistency yields the superpentagon equations, and more details are in App.~\ref{App:pentagon_anomaly}.
\end{itemize}


\paragraph{superstrip algebra}

For a general 2D bosonic theory with a finite global symmetry described by a fusion category $\mathscr{C}$, the IR phase of the symmetry is captured by a $\mathscr{C}$-module category $\mathcal{M}$ \cite{Fuchs:2012dt}, which can be thought as the generalization of ``representation" of the fusion category $\mathscr{C}$. In the case of a gapped phase, the simple objects $v_i$ in $\mathcal{M}$ represent vacua labeled as $|i\rangle$. In what follows, we assume that the gapped phase of a 2D fermionic theory can still be described by a module $\mathcal{M}$ of the superfusion category $\mathscr{C}$ with the fusion rule
    \begin{equation}
        \mathcal{L}_a \cdot v_i = \sum_j\widetilde{N}_{ai}^{j,\bf b} v_j + \sum_j\widetilde{N}_{ai}^{j,\bf f} v_j \,,
    \end{equation}
where $\widetilde{N}_{ai}^{j,\bf b(\bf f)}$ are non-negative integers reflecting the dimension of the junction vector space
    \begin{equation}
        V_{\mathcal{L}_a,v_i,\overline{v}_j} = \mathbb{C}^{\widetilde{N}_{ai}^{j,\bf b}|\widetilde{N}_{ai}^{j,\bf f}}\,,
    \end{equation}
which is depicted as
    \begin{equation}
        \begin{gathered}
     \begin{tikzpicture}[scale=0.8]
        \filldraw[gray,opacity=0.3] (-2.5,0)--(-3.5,0)--(-3.5,2)--(-2.5,2)--(-2.5,0);
        \draw[ultra thick,MidnightBlue,->-=.55] (-2.5,2)--(-2.5,1);
        \draw[ultra thick,MidnightBlue,->-=.55] (-2.5,1)--(-2.5,0);
        \draw[densely dashed,ultra thick,BurntOrange,->-=.6] (-1.25,1)--(-2.5,1);
        \node at (-2.85,1) {$\textcolor{black}\beta$};
         \filldraw[black] (-2.5,1) circle (2pt);
        \node at (-2.9,0.3) {$v_j$};
        \node at (-2.9,1.7) {$v_i$};
        \node at (-.75,1) {$\mathcal{L}_a$};
    \end{tikzpicture}    
        \end{gathered}\quad {\color{black}{\beta}} = ( {\color{black}{\beta^{\bf b}}},\,{\color{black}{\beta^{\bf f}}} )
    \end{equation}
with $\beta^{\bf b}=1,\cdots,\widetilde{N}_{ai}^{j,\bf b}$ and $\beta^{\bf f}=1,\cdots,\widetilde{N}_{ai}^{j,\bf f}$. We use solid lines and dashed lines to represent vacua and symmetry operators separately.

The Hilbert space of the theory can be defined by specifying the vacua at spatial infinities. For instance, $\mathcal{H}_{i,j}$ denotes 
the Hilbert space where the vacuum at $\sigma \rightarrow - \infty$ 
is $|i\rangle$ while $|j\rangle$ at $\sigma \rightarrow+\infty$. If $|i\rangle=|j\rangle$ the Hilbert space $\mathcal{H}_{i,i}$ consists of the particle excitations on the vacuum $|i\rangle$. On the other hand, if $|i\rangle\neq |j\rangle$, $\mathcal{H}_{i,j}$ contains the solitonic states interpolating between two vacua. We introduce the total Hilbert space as $\mathcal{H}=\oplus_{i,j} \mathcal{H}_{i,j}$.

Non-invertible symmetries map states between different sectors, constraining possible spectrum. However, their non-invertible nature --- i.e. the presence of non-trivial (co)kernels --- means these maps are not simple bijections, which creates ambiguities in the naive particle/soliton multiplet structure. The ``strip algebra" $\textbf{Str}_{\mathscr{C}}(\mathcal{M})$, defined by input symmetry category $\mathscr C$ and boundary module $\mathcal M$, was introduced for bosonic systems \cite{Cordova:2024vsq, Cordova:2024iti} to resolve this. It provides a systematic formalism to project onto the correct multiplet structure and organize the true state degeneracies consistent with the symmetry $\mathscr C$.


In this letter, we generalize this to fermionic systems, introducing the ``superstrip algebra" $\textbf{sStr}_{\mathscr{C}}(\mathcal{M})$ in order to incorporate the aforementioned fermionic nature of TDLs and junctions. The basis in $\textbf{sStr}_{\mathscr{C}}(\mathcal{M})$ takes the form
\begin{equation}\label{fig:superstrip}
\begin{gathered}
     \begin{tikzpicture}[scale=0.8]
        \filldraw[gray,opacity=0.3] (0,0)--(1,0)--(1,2)--(0,2)--(0,0);
        \filldraw[gray,opacity=0.3] (2.5,0)--(3.5,0)--(3.5,2)--(2.5,2)--(2.5,0);
        \draw[ultra thick,MidnightBlue,-<-=.55] (1,0)--(1,1);
        \draw[ultra thick,MidnightBlue,-<-=.55] (1,1)--(1,2);
        \draw[ultra thick,MidnightBlue,->-=.65] (2.5,0)--(2.5,1);
        \draw[ultra thick,MidnightBlue,->-=.7] (2.5,1)--(2.5,2);
        \draw[densely dashed,ultra thick,BurntOrange,-<-=.55] (1,0.9)--(2.5,1.15);
        \node at (0.7,.9) {$\textcolor{black}\alpha$};
        \node at (2.8,1.15) {$\textcolor{black}\beta$};
         \filldraw[black] (1,0.9) circle (2pt);
         \filldraw[black] (2.5,1.15) circle (2pt);
        \node at (1.25,0.25) {$i$};
        \node at (1.25,1.75) {$k$};
        \node at (2.25,0.25) {$j$};
        \node at (2.25,1.75) {$l$};
        \node at (1.75,0.75) {$a$};
    \end{tikzpicture}
\end{gathered}\ .
\end{equation}
The subscripts $a$ and $i$ refer to 
the TDL ${\cal L}_a$ and the vacuum $|i\rangle$. 
Physically, it defines a map $a: \mathcal H_{i,j}\rightarrow\mathcal H_{k,l}$. Our convention is that the possible left dot $\alpha$ is lower than the possible right dot $\beta$. When this order is reversed, an extra sign factor $(-1)^{s(\alpha) s(\beta)}$ appears,  where $s(\alpha)=0,1$ denotes the $\mathbb{Z}_2$-grading of the junction $\alpha$.

A generic element is a linear combination over complex numbers. The multiplication of two elements is given by concatenating one element on top of the other, as depicted below: 
\begin{align}
\begin{gathered}
        \begin{tikzpicture}[scale=0.8]
        \filldraw[gray,opacity=0.3] (-0.5,-1)--(0,-1)--(0,1)--(-0.5,1)--(-0.5,-1);
        \filldraw[gray,opacity=0.3] (1,-1)--(1.5,-1)--(1.5,1)--(1,1)--(1,-1);
            \draw[ultra thick,MidnightBlue,->-=.55] (0,1)--(0,0);
            \draw[ultra thick,MidnightBlue,->-=.65] (0,0)--(0,-1);
            \draw[ultra thick,MidnightBlue,->-=.75] (1,0)--(1,1);
            \draw[ultra thick,MidnightBlue,->-=.65] (1,-1)--(1,0);
            \draw[densely dashed,ultra thick,BurntOrange,-<-=.6] (0,0)--(1,0.15);
            \node at (0,1.25) {$i$};
            \node at (1,1.25) {$j$};
            \node at (0,-1.25) {$k$};
            \node at (1,-1.25) {$l$};
            \node at (-0.25,0.05) {$\textcolor{black}\alpha$};
            \node at (1.25,0.15) {$\textcolor{black}\beta$};
            \node at (0.5,0.35) {$a$};
            \filldraw[black] (0,0) circle (1.5pt);
            \filldraw[black] (1,0.15) circle (1.5pt);
        \end{tikzpicture}
    \end{gathered}  \otimes
    \begin{gathered}
        \begin{tikzpicture}[scale=0.8]
        \filldraw[gray,opacity=0.3] (-0.5,-1)--(0,-1)--(0,1)--(-0.5,1)--(-0.5,-1);
        \filldraw[gray,opacity=0.3] (1,-1)--(1.5,-1)--(1.5,1)--(1,1)--(1,-1);
            \draw[ultra thick,MidnightBlue,->-=.55] (0,1)--(0,0);
            \draw[ultra thick,MidnightBlue,->-=.65] (0,0)--(0,-1);
            \draw[ultra thick,MidnightBlue,->-=.75] (1,0)--(1,1);
            \draw[ultra thick,MidnightBlue,->-=.65] (1,-1)--(1,0);
            \draw[densely dashed,ultra thick,BurntOrange,-<-=.6] (0,0)--(1,0.15);
            \node at (0,1.25) {$r$};
            \node at (1,1.25) {$s$};
            \node at (0,-1.25) {$m$};
            \node at (1,-1.25) {$n$};
            \node at (-0.25,0.05) {$\textcolor{black}\gamma$};
            \node at (1.25,0.15) {$\textcolor{black}\delta$};
            \node at (0.5,0.4) {$b$};
            \filldraw[black] (0,0) circle (1.5pt);
            \filldraw[black] (1,0.15) circle (1.5pt);
        \end{tikzpicture}
    \end{gathered}\, 
 \mapsto \, \delta_{k,r} \delta_{l,s}\,
\begin{gathered}
            \begin{tikzpicture}[scale=0.85]
            \filldraw[gray,opacity=0.3] (-0.5,-1)--(0,-1)--(0,1)--(-0.5,1)--(-0.5,-1);
        \filldraw[gray,opacity=0.3] (1,-1)--(1.5,-1)--(1.5,1)--(1,1)--(1,-1);
                \draw[ultra thick,MidnightBlue,->-=.55] (0,1)--(0,0.4);
                \draw[ultra thick,MidnightBlue,->-=.65] (0,0.4)--(0,-0.4);
                \draw[ultra thick,MidnightBlue,->-=.85] (0,-0.4)--(0,-1);
                \draw[ultra thick,MidnightBlue,-<-=.55] (1,1)--(1,0.4);
                \draw[ultra thick,MidnightBlue,-<-=.55] (1,0.4)--(1,-0.4);
                \draw[ultra thick,MidnightBlue,-<-=.75] (1,-0.4)--(1,-1);
                \draw[densely dashed,ultra thick,BurntOrange,-<-=.6] (0,0.4)--(1,0.55);
                \draw[densely dashed,ultra thick,BurntOrange,-<-=.6] (0,-0.4)--(1,-0.25);
                \node at (0,1.25) {$i$};
                \node at (1,1.25) {$j$};
                \node at (1.20,0.2) {$l$};
                \node at (-0.25,0) {$k$};
                \node at (0,-1.25) {$m$};
                \node at (1,-1.25) {$n$};
                \node at (0.5,0.75) {$a$};
                \node at (0.5,-0.0) {$b$};
                \node at (1.25,0.70) {$\textcolor{black}\beta$};
                \node at (1.20,-0.35) {$\textcolor{black}\delta$};
                \node at (-0.25,0.5) {$\textcolor{black}\alpha$};
                \node at (-0.25,-0.55) {$\textcolor{black}\gamma$};
            \filldraw[black](0,0.4) circle (1.5pt);
            \filldraw[black] (1,0.55) circle (1.5pt);
            \filldraw[black] (0,-0.4) circle (1.5pt);
            \filldraw[black] (1,-0.25) circle (1.5pt);
        \end{tikzpicture}
    \end{gathered}
        \end{align}
In fact, the superstrip algebra is a $C^\ast$-weak Hopf superalgebra, and details regarding its mathematical structure will be discussed in the forthcoming paper \cite{toappear}.

In this paper, we focus on the case where $\mathcal{M}=\mathscr{C}$ as its regular module, which describes the gapped spontaneous symmetry breaking (SSB) phase. In this case, all vacua can be obtained from the identity vacuum $|I\,\rangle$ by acting $\mathcal L_i$,
\begin{align}
    \mathcal L_i|I\,\rangle=|i\,\rangle\,,
\end{align}
and thus are labeled by these simple TDLs. We emphasize that the fermion parity $(-1)^F$ is assumed to be non-anomalous and unbroken, so $\mathscr{C}$ is actually a reduced superfusion category by factorizing out $(-1)^F$. One immediate consequence that follows, as shown in App.~\ref{app:Fermionization}, is the correlation between the spin-statistics of Ramond vacuum $|i\rangle$ and the type of $\mathcal{L}_i$:
\begin{itemize}
    \item $m$-type objects lead to bosonic vacua, the same as in the fusion category case;
    \item $q$-type objects hosting a 1D Majorana fermion necessarily give rise to fermionic vacua.
\end{itemize}

Moreover, irreducible representations of the superstrip algebra $\textbf{sStr}_{\mathscr{C}}(\mathscr{C})$ are also uniquely labeled by the simple TDLs, generalizing the bosonic case studied in \cite{Cordova:2024iti}.
Thus, a quiver diagram can conveniently encode 
various properties of an irreducible representation associated with $\mathcal{L}_a$:  
\begin{itemize}
    \item For each simple TDL in $\mathscr{C}$, we draw a node.
    \item Given any two nodes $i$ and $j$, we draw $N_{ai}^{j,\bf b}$ blue arrows and $N_{ai}^{j,\bf f}$ red arrows from $i$ to $j$.
\end{itemize}
The quiver diagram visualizes the soliton spectrum: blue arrows denote bosonic excitations and red arrows denote fermionic excitations in the corresponding Hilbert space. Furthermore, \cite{Cordova:2024vsq} shows that a theory with trivial one-form symmetry is equivalent to having a connected quiver diagram for the complete set of stable solitons.

\section{Tri-critical Ising model with the least relevant deformations} 
\begin{figure}
    \begin{tikzpicture}
        \node (top_left) at (0,3) {\color{Teal} Ising Model};
        \node (middle_left) at (0,1.5) {\color{Teal} TIM};
        \node (bottom_left) at (0,0) {\color{Teal} Ising SSB};
        \node (top_right) at (6,3) {\color{Teal} Majorana fermion};
        \node (middle_right) at (6,1.5) {\color{Teal} $\mathcal{N}=1$ $A_2$ MM};
        \node (bottom_right) at (6,0) {\color{Teal} $\mathbb{Z}_2$ SSB};
        \draw[sharp arrow, thick] (middle_left) -- (top_left);
        \draw[sharp arrow, thick] (middle_left) -- (bottom_left);
        \draw[sharp arrow, thick] (middle_right) -- (top_right);
        \draw[sharp arrow, thick] (middle_right) -- (bottom_right);
        \node at (-0.5,2.15) {$\lambda>0$};
        \node at (-0.5,0.85) {$\lambda<0$};
        \node at (6.5,2.15) {$\lambda>0$};
        \node at (6.5,0.85) {$\lambda<0$};
        \draw[sharp arrow, thick] ([yshift=2pt]top_left.east) -- ([yshift=2pt]top_right.west) node[midway,above] {fermionize};
        \draw[sharp arrow, thick] ([yshift=-2pt]top_right.west) -- ([yshift=-2pt]top_left.east) node[midway,below] {gauge $(-1)^F$};
        \draw[sharp arrow, thick] ([yshift=2pt]middle_left.east) -- ([yshift=2pt]middle_right.west) node[midway,above] {fermionize};
        \draw[sharp arrow, thick] ([yshift=-2pt]middle_right.west) -- ([yshift=-2pt]middle_left.east) node[midway,below] {gauge $(-1)^F$};
        \draw[sharp arrow, thick] ([yshift=2pt]bottom_left.east) -- ([yshift=2pt]bottom_right.west) node[midway,above] {fermionize};
        \draw[sharp arrow, thick] ([yshift=-2pt]bottom_right.west) -- ([yshift=-2pt]bottom_left.east) node[midway,below] {gauge $(-1)^F$};
    \end{tikzpicture}
    \caption{The duality between TIM and $\mathcal{N}=1$ $A_2$ minimal model (MM). The two columns are related by fermionization and bosonization.}
    \label{Tri-critical duality}
\end{figure}
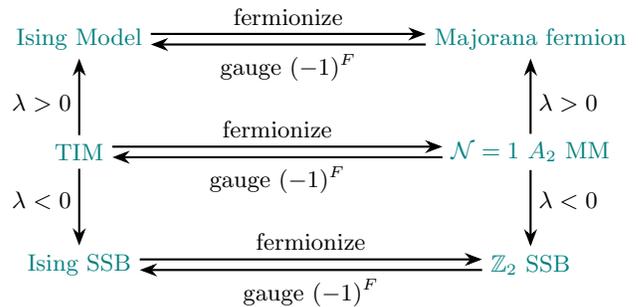

Before diving into the world of $\mathcal{N} = 2$ minimal models, let us first tackle a concrete but simple model to motivate the study of superstrip algebra. An example is the tri-critical Ising model (TIM), perturbed by the least relevant term $\lambda \int \phi_{\frac{3}{5},\frac{3}{5}}$. 
For $\lambda >0$, the theory flows to the gapless Ising model, whereas for $\lambda <0$ it flows to a gapped phase with three vacua where the Ising symmetry is spontaneously broken.
As depicted in Fig.~\ref{Tri-critical duality}, the TIM has hidden supersymmetry (SUSY) and can be fermionized to the $\mathcal{N}=1$ $A_2$ minimal model.
%

The TIM under the $\phi_{\frac{3}{5},\frac{3}{5}}$ deformation allows the ${\cal N}=1$ supersymmetric 
Landau–Ginzburg (LG) description with the superpotential 
$\mathcal{W}(\Phi) =\frac{1}{3} \Phi^3 + \lambda \Phi$ 
, where $\Phi$ is an $\mathcal{N}=1$ superfield, $\Phi = \phi + \bar{\theta} \psi + \bar{\theta} \theta F$. The potential term then becomes 
\begin{equation}\label{potential-N1}
    V = \frac{1}{2}|\mathcal{W}(\phi)'|^2 + \frac{1}{2}\mathcal{W}(\phi)''\bar{\psi} \psi= \frac{1}{8} (\phi^2 + 2\lambda)^2 + \phi \bar{\psi} \psi \,. \nonumber
\end{equation} 
The LG model has a $\mathbb{Z}^F_2 \times \mathbb{Z}_2^\eta$ global symmetry,  
the former corresponding to the fermion parity $(-1)^F$ and the latter to
the chiral $R$-symmetry 
\begin{equation}
        \eta:\quad\phi \rightarrow -\phi\,,\quad \psi\rightarrow \gamma_5 \psi\,.
        \label{eq:N=1_F_L}
\end{equation}
%
For $\lambda<0$, the potential $V$ has two minima at $\phi = \pm \sqrt{2}|\lambda|$, where the $\mathbb{Z}_2^\eta$ symmetry is spontaneously broken. At each minimum, both the scalar and fermion become massive so that the theory flows to a gapped phase. The two minima
give rise to the degenerate vacua, denoted by $|I\rangle$ and $|\eta\rangle$, which are interchanged under \eqref{eq:N=1_F_L}
	\begin{equation}
	\eta |I\rangle = |\eta\rangle\,, \quad \eta |\eta\rangle = |I\rangle\,.
	\end{equation}
The fermion parity remains unbroken. The two vacua differ by the sign of mass $m=\langle \phi \rangle$ of the Majorana fermion. One can thus
designate, without loss of generality, $|I\rangle$ as the trivial phase and $|\eta\rangle$ as the non-trivial fermionic symmetry protected topological (SPT) phase \cite{Witten:2023snr}. 

Since the $\mathbb{Z}_2^\eta$ symmetry is spontaneously broken, the $\eta$-line along the temporal direction 
can be viewed as a soliton interpolating between two vacua. One can argue that, in the presence of the soliton, 
the fermion field $\psi$ has a single real zero mode localized at the center of the soliton. As a 
consequence, $\eta$-line has to be dressed by a 1D Majorana fermion and thus becomes a 
$q$-type TDL obeying the fusion rule below, 
\begin{equation}
    \eta \times \eta = \mathbb C^{1|1}I\,,\quad (N_{\eta \eta}^{I,\bf b}=N_{\eta \eta}^{I,\bf f}=1)
\end{equation}
where $V_{\eta,\eta,I}=\mathbb{C}^{1|1}$, and we will use $\alpha=0,1$ to label the two channels $\mathbb{C}^{1|0}$ and $\mathbb{C}^{0|1}$. Physically, the two Majorana zero modes tied with a pair of $\eta$-lines combine into a Dirac zero mode, which spans a 2-dimensional superspace upon quantization. We denote the corresponding superfusion category as $\mathscr {C}_q^0$.

\begin{figure}[H]
\centering
    \begin{tikzpicture}[
    dot_style/.style={circle, fill, inner sep=1.5pt},
    blue_arc/.style={Cerulean, thick, -Stealth, bend left=25},
    red_arc/.style={WildStrawberry, thick, -Stealth, bend left=25},
    label_below/.style={below=2pt}, scale=1
]

\node at (-1.5, 0) {$\mathcal R_{\eta}$:};

    \node[dot_style, label={[label distance=3pt]left:$|I\rangle$}] (n0) at (0, 0) {};
    \coordinate (n0pr) at (0.05, .25) {};
    \coordinate (n0pr2) at (0.1, .08) {};

    \coordinate (n0nr2) at (0.1, -.08) {};
    \coordinate (n0nr) at (0.1, -.25) {};

    \node[dot_style, label={[label distance=3pt]right:$|\eta\rangle$}] (n1) at (3, 0) {};
   
    \coordinate (n1pl) at (2.95, .25) {};
    \coordinate (n1pl2) at (2.9, .08) {};
    
    \coordinate (n1nl2) at (2.9, -.08) {};
    \coordinate (n1nl) at (2.9, -.25) {};

    \draw[blue_arc] (n0pr) to (n1pl);
    \draw[red_arc] (n0pr2) to (n1pl2);

    \draw[blue_arc] (n1nl2) to (n0nr2);
    \draw[red_arc] (n1nl) to (n0nr);
\end{tikzpicture}
\caption{The quiver diagram of the representation $\mathcal R_{\eta}$.}
\label{fig:Cq0quiver}
\end{figure}
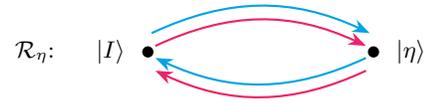

The superstrip algebra $\textbf{sStr}_{\mathscr {C}_q^0}(\mathscr {C}_q^0)$ consists of the building block \eqref{fig:superstrip} with $i,j,k,l,a \in \{I,\eta \}$.
Begin with a bosonic soliton $|K_{\eta I}\rangle \in \mathcal{H}_{\eta,I}$, it can be mapped to a fermionic soliton $\widetilde{|K_{\eta I}\rangle}$ via
\begin{equation}
        \begin{gathered}
     \begin{tikzpicture}[scale=1]
        \filldraw[gray,opacity=0.3] (0,0)--(1,0)--(1,2)--(0,2)--(0,0);
        \filldraw[gray,opacity=0.3] (2.5,0)--(3.5,0)--(3.5,2)--(2.5,2)--(2.5,0);
        \filldraw[Cerulean,opacity=0.1] (1,0)--(1,0.9)--(2.5,1.1)--(2.5,0)--(1,0);
        \filldraw[WildStrawberry,opacity=0.1] (1,0.9)--(2.5,1.1)--(2.5,2)--(1,2)--(1,0.9);
        \draw[ultra thick,Cerulean,-<-=.55] (1,0)--(1,.9);
        \draw[ultra thick,WildStrawberry,-<-=.55] (1,.9)--(1,2);
        \draw[ultra thick,Cerulean,->-=.55] (2.5,0)--(2.5,1.1);
        \draw[ultra thick,WildStrawberry,->-=.55] (2.5,1.1)--(2.5,2);
        \draw[densely dashed,ultra thick,BurntOrange,-<-=.55] (1,0.9)--(2.5,1.1);
        \node at (0.5,.9) {\footnotesize$\color{WildStrawberry}\alpha=1$};
        \node at (3,1.1) {\footnotesize$\color{Cerulean}\beta=0$};
        \filldraw[WildStrawberry] (1,0.9) circle (2pt);
        \filldraw[Cerulean] (2.5,1.1) circle (2pt);
        \node at (1., -.2) {\footnotesize$|\eta\rangle$};
        \node at (1.,2.2) {\footnotesize $|\eta\rangle$};
        \node at (2.5,-.2) {\footnotesize$|I\rangle$};
        \node at (2.5,2.2) {\footnotesize $|I\rangle$};
        \node at (1.75,0.65) {$I$};
    \end{tikzpicture}        
    \end{gathered}
\end{equation}
which reverses the statistics of the soliton due to the fermionic junction $\alpha=1$. The blue and red shaded regions represent the bosonic and fermionic excitations, respectively. It can also be mapped to an anti-soliton in $\mathcal{H}_{I,\eta}$ through two inequivalent channels
\begin{equation}
\begin{gathered}
     \begin{tikzpicture}[scale=1]
        \filldraw[gray,opacity=0.3] (0,0)--(1,0)--(1,2)--(0,2)--(0,0);
        \filldraw[gray,opacity=0.3] (2.5,0)--(3.5,0)--(3.5,2)--(2.5,2)--(2.5,0);
        \filldraw[Cerulean,opacity=0.1] (1,0)--(1,0.9)--(2.5,1.1)--(2.5,0)--(1,0);
        \filldraw[Cerulean,opacity=0.1] (1,0.9)--(2.5,1.1)--(2.5,2)--(1,2)--(1,0.9);
        \draw[ultra thick,Cerulean,-<-=.55] (1,0)--(1,.9);
        \draw[ultra thick,Cerulean,-<-=.55] (1,.9)--(1,2);
        \draw[ultra thick,Cerulean,->-=.55] (2.5,0)--(2.5,1.1);
        \draw[ultra thick,Cerulean,->-=.55] (2.5,1.1)--(2.5,2);
        \draw[densely dashed,ultra thick,BurntOrange,-<-=.55] (1,0.9)--(2.5,1.1);
        \node at (0.5,.9) {\footnotesize$\color{Cerulean}\alpha=0$};
        \node at (3,1.1) {\footnotesize$\color{Cerulean}\beta=0$};
        \filldraw[Cerulean] (1,0.9) circle (2pt);
        \filldraw[Cerulean] (2.5,1.1) circle (2pt);
        \node at (1., -.2) {\footnotesize$|\eta\rangle$};
        \node at (1.,2.2) {\footnotesize $|I\rangle$};
        \node at (2.5,-.2) {\footnotesize$|I\rangle$};
        \node at (2.5,2.2) {\footnotesize $|\eta\rangle$};
        \node at (1.75,0.65) {$\eta$};
    \end{tikzpicture}   
\end{gathered}\quad   \begin{gathered}
     \begin{tikzpicture}[scale=1]
        \filldraw[gray,opacity=0.3] (0,0)--(1,0)--(1,2)--(0,2)--(0,0);
        \filldraw[gray,opacity=0.3] (2.5,0)--(3.5,0)--(3.5,2)--(2.5,2)--(2.5,0);
        \filldraw[Cerulean,opacity=0.1] (1,0)--(1,0.9)--(2.5,1.1)--(2.5,0)--(1,0);
        \filldraw[WildStrawberry,opacity=0.1] (1,0.9)--(2.5,1.1)--(2.5,2)--(1,2)--(1,0.9);
        \draw[ultra thick,Cerulean,-<-=.55] (1,0)--(1,.9);
        \draw[ultra thick,WildStrawberry,-<-=.55] (1,.9)--(1,2);
        \draw[ultra thick,Cerulean,->-=.55] (2.5,0)--(2.5,1.1);
        \draw[ultra thick,WildStrawberry,->-=.55] (2.5,1.1)--(2.5,2);
        \draw[densely dashed,ultra thick,BurntOrange,-<-=.55] (1,0.9)--(2.5,1.1);
        \node at (0.5,.9) {\footnotesize$\color{WildStrawberry}\alpha=1$};
        \node at (3,1.1) {\footnotesize$\color{Cerulean}\beta=0$};
        \filldraw[WildStrawberry] (1,0.9) circle (2pt);
        \filldraw[Cerulean] (2.5,1.1) circle (2pt);
        \node at (1., -.2) {\footnotesize$|\eta\rangle$};
        \node at (1.,2.2) {\footnotesize $|I\rangle$};
        \node at (2.5,-.2) {\footnotesize$|I\rangle$};
        \node at (2.5,2.2) {\footnotesize $|\eta\rangle$};
        \node at (1.75,0.65) {$\eta$};
    \end{tikzpicture} 
\end{gathered}
\end{equation}
Here we set $\beta=0$, since the red dot on the $\beta$ junction can move along the $\eta$ line to the $\alpha$ junction. As a consequence, the irreducible representation of solitons consists of four states, two solitons $|K_{\eta I}\rangle,\widetilde{|K_{\eta I}\rangle}$ and two anti-solitons $|K_{I\eta}\rangle,\widetilde{|K_{I\eta}\rangle}$, with opposite statistics in each pair, represented by the quiver diagram in Fig.~\ref{fig:Cq0quiver}. It agrees perfectly with the integrability results \cite{Schoutens:1990vb, Ahn:1990gn}.

As a final remark, there have been some debates regarding the representation of solitons. 
Since a soliton supports only a single Majorana zero mode
, one might expect the soliton $|K_{\eta I}\rangle$ should be one-dimensional and invariant under the action of the broken supercharge $Q_2 |K_{\eta I}\rangle \sim |K_{\eta I}\rangle$, rather than being transformed into $\widetilde{|K_{\eta I}\rangle}$. This viewpoint was accepted in 
\cite{Ritz:2000xa,Losev:2000mm, Losev:2001uc}. However, the fermion parity $(-1)^F$ ceases to be a symmetry due to $\{(-1)^F,Q_2\}=0$. In contrast, within our framework, we retain $(-1)^F$ as an intrinsic element of the superfusion category. Then $ Q_2 |K_{\eta I}\rangle \sim \widetilde{|K_{\eta I}\rangle}$ and the soliton is doubly degenerate \cite{Witten:1977xv, note:degeneracy}. 


\section{$\mathbf{ADE}$ classification of $\mathcal N=2$ minimal integrable models}
\label{sec:ADE}

We now analyze the $\mathcal N=2$ $A_{k+1}$ minimal models perturbed by the least relevant chiral operator $\Phi_{(k,k)}$. First, it is known that the model has $k$ relevant deformations preserving the supersymmetry. One can describe $(k+1)$-th minimal model by an LG model with superpotential ${\cal W}(\Phi)= \Phi^{k+2}/(k+2)$ where $\Phi$ is an ${\cal N}=2$ chiral superfield. Moreover, the generic relevant deformation can be described 
in the LG model as follows,  
\begin{align}
    {\cal W}_\text{def}(\Phi) = \frac{1}{k+2} \Phi^{k+2} 
    + \sum_{j=2}^{k+1} \frac{\nu_j}{k+2-j} \Phi^{k+2-j}\ , \label{N2defW}
\end{align}
where $\nu_j$ are deformation parameters. In this paper we only consider those $\nu_j$ such that the deformed theory becomes massive and has $(k+1)$ distinct vacua, determined by ${\cal W}_\text{def}^{\,\prime}(\Phi^{(r)})=0$ ($r=1,2,..,k+1$). Since the Witten index of the model is given by $(k+1)$ independent of $\nu_j$, 
one can argue that all SUSY vacua are bosonic. 

One can ask how many solitons preserving half of the SUSY, often 
referred to as the Bogomolny–Prasad–Sommerfield (BPS) states, exist for a given pair of $(k+1)$ vacua. In fact, it was shown in \cite{Cecotti:1992rm} that such a BPS spectrum is piecewise constant over the parameter space $\{\nu_j\}$, and exhibits discontinuities across the walls of marginal stability. 

We are interested in the BPS spectrum of the minimal model with the least relevant deformation, denoted by $A_{k+1}^{\rm def}$. 
The models $A_{k+1}^{\rm def}$ are known to be integrable, and allow us to compute their exact S-matrices and BPS spectrum therein. 
More precisely, the model $A_{k+1}^{\rm def}$ possesses $k$ soliton (anti-soliton) supermultiplets of equal mass \cite{Bernard:1990ti}. 

In terms of the LG model, the least relevant deformation corresponds to the theory at a special point on the parameter space where the superpotential ${\cal W}_\text{def}(\Phi)$ becomes the Chebyshev polynomial \cite{Dijkgraaf:1990dj}. The SUSY vacua $\Phi^{(r)}$ are then aligned linearly in the $\Phi$-plane. Then indeed one can show that the (anti-)solitons exist only between adjacent vacua $r$ and $(r+1)$, and have equal mass determined by $m_{\text{BPS}}=\left|\mathcal W(\Phi_{\rm def}^{(r)})-\mathcal W(\Phi_{\rm def}^{(r+1)})\right|$ \cite{Fendley:1991ve}.

\subsection{Superfusion category in $A_{k+1}^{\rm def}$}

\begin{figure}
    \centering
\begin{tikzpicture}[
    node distance=3.5cm and 4.5cm,
    every node/.style={align=center}, 
    arrow_style/.style={-latex, thick, Teal}, 
    text_style/.style={Teal}
]

\node (top_left) at (0, 2.5) {$\color{Teal}\dfrac{\mathrm{SU}(2)_k \times \mathrm{U}(1)_2}{\mathrm{U}(1)_{k+2}}$};
\node (top_right) at (6, 2.5) {$\color{Teal}\mathcal{N}=2 \ \ A_{k+1}$};
\node (bottom_left) at (0, 0) {$\color{Teal}\mathcal{B}\big[A_{k+1}^{\text{def}}\big]$};
\node (bottom_right) at (6, 0) {$\color{Teal}\mathcal{N}=2 \ \ A_{k+1}^{\text{def}}$};

\draw[sharp arrow, thick] (top_left) -- (bottom_left); 
\node at (-.5,1.5) {$\Phi_{(k,k)}^{\rm top}$};
\node at (-.6,1.167) {\rm deform.};
\draw[sharp arrow, thick] (top_right) -- (bottom_right); 
\node at (5.5,1.5) {$\Phi_{(k,k)}$};
\node at (5.4,1.167) {\rm deform.};

\draw[sharp arrow, thick] ([yshift=2pt]top_left.east) -- ([yshift=2pt]top_right.west) node[midway, above] {fermionize $\mathcal{L}_{(k,k+2)}$};
\draw[sharp arrow, thick]  ([yshift=-2pt]top_right.west) -- ([yshift=-2pt]top_left.east)  node[midway, below] {gauge $(-1)^F$};

\draw[sharp arrow, thick] ([yshift=2pt]bottom_left.east) -- ([yshift=2pt]bottom_right.west) node[midway, above] {fermionize $\mathcal{L}_{(k,k+2)}$};
\draw[sharp arrow, thick] ([yshift=-2pt]bottom_right.west) -- ([yshift=-2pt]bottom_left.east) node[midway, below] {gauge $(-1)^F$};

\node[font=\large, text_style] (c) at (0, -.917) {${\tilde{\mathscr{C}}_b}$};
\node[font=\large, text_style] (cf) at (6.1, -.917) {$\tilde{\mathscr C}_{A_{k+1}}$};

\node at (0, -.417) {$\bigcup$};
\node at (6, -.417) {$\bigcup$};

\draw[sharp arrow, thick, Teal] (c) -- (cf) node[midway, below] {fermionic condensation of $\mathcal{L}_{(k,k+2)}$};

\end{tikzpicture}
    \caption{Commutative diagram to compute the preserved superfusion category $\tilde{\mathscr C}_{A_{k+1}}$ in $A_{k+1}^{\rm def}$. }
    \label{fig:CD}
\end{figure}
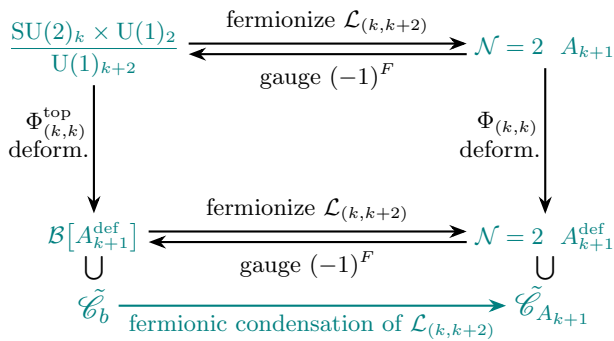

Let us discuss the superfusion category of $A_{k+1}^{\rm def}$, denoted by $\tilde{\mathscr C}_{A_{k+1}}$ in what follows. Here, the tilde indicates that the category is associated with the deformed theory. 
As explained above, the least relevant deformation is triggered 
by adding the superspace operator $\Phi_{(k,k)}$ via its 
top component $\Phi_{(k,k)}^{\rm top}$: 
$\lambda\int{\rm d}^2x\,{\rm d}^2\theta\,\Phi_{(k,k)}=\lambda \int {\rm d}^2x\ \Phi_{(k,k)}^{\rm top}\,$.
The key feature of $\Phi_{(k,k)}^{\rm top}$ is that, 
although it is an $\mathcal N=2$ superconformal descendant, 
it is a primary of the bosonic sub-algebra. This observation allows us to construct $\tilde{\mathscr C}_{A_{k+1}}$ in following two steps, 
summarized in Fig.~\ref{fig:CD}.
\noindent
First, we compute the preserved fusion category $\mathscr C_b$ in the bosonized $A^{\rm def}_{k+1}$ model, denoted $\mathcal{B}[A_{k+1}^{\text{def}}]$. Second, we perform fermionic condensation of $\mathscr C_b$ \cite{Zhou:2021ulc} to obtain $\tilde{\mathscr C}_{A_{k+1}}$. As a result, one obtains the category $\tilde{\mathscr C}_{A_{k+1}}=\{\,\mathcal L_a\,|a=0,1,2,\dots k\}$ 
with fusion rules below \eqref{eq: su_2_N=2}
\begin{equation}
    \mathcal L_a\cdot \mathcal L_{a'}=
    \begin{cases}
        \mathbb C^{1|0}\sum_{\alpha=|a-a'|,\,{\rm step \ 2}}^{\min(a+a',\, 2k-a-a')}  \mathcal L_\alpha\,,\quad {\rm if}\ \ a\cdot a'\ {\rm even}\,, \\[5pt]
        \mathbb C^{0|1}\sum_{\alpha=|a-a'|,\,{\rm step \ 2}}^{\min(a+a',\, 2k-a-a')}  \mathcal L_\alpha\,,\quad {\rm if}\ \ a\cdot a'\ {\rm odd}\,,
    \end{cases}
    \label{eq: su_2_N=2}
\end{equation}
where we sum over indices of the same parity. The above category can be identified as a \emph{fermionic} variant of ${\rm SU}(2)_k$, denoted by ${\rm SU}(2)_k^{\mathcal N=2}$.
The readers can find the technical details in App.~\ref{app:Fermionization}. 


\subsection{Vacua, Witten Index, and 't Hooft anomalies}
\label{sec:section_4.B}

The $(k+1)$ vacua of the $A_{k+1}^{\rm def}$ model completely spontaneously break the preserved superfusion categorical symmetry $\tilde{\mathscr C}_{A_{k+1}}$ that contains $(k+1)$ simple objects.  
%
In other words, the vacua form the regular module of the category $\mathcal{M}_{\tilde{\mathscr C}_{A_{k+1}}}\simeq \tilde{\mathscr C}_{A_{k+1}}$.

Furthermore, eq. \eqref{eq:Z2_on_C} shows that all objects in $\tilde{\mathscr C}_{A_{k+1}}$ are of $m$-type. As explained in App.~\ref{app:Fermionization}, this implies that all $(k+1)$ vacua should be bosonic, and therefore the Witten index becomes
$(k+1)$, i,e, $I^{\mathcal{N}=2}_W(A_{k+1}^{\textrm{def.}})=|\tilde{\mathscr C}_{A_{k+1}}|=k+1$. This result perfectly matches the $\mathcal N=2$ supersymmetry calculation \cite{Lerche:1989uy}, yet here it follows purely from a \emph{generalized symmetry} perspective.

An important special case arises for $A_{2}^{\rm def}$, where the superpotential $\mathcal{W}(\Phi) =\frac{1}{3} \Phi^3 + \lambda \Phi$ takes the same form as its $\mathcal{N}=1$ counterpart. The difference is that the $\mathcal{N}=2$ superfield $\Phi$ has a complex scalar field $\phi$ and a Dirac field $\psi$. The non-trivial element of the preserved symmetry category $\tilde{\mathscr C}_{A_2}={\rm SU}(2)_1^{\mathcal N=2}$ is again the $\mathbb Z_2$-chiral symmetry line $\eta\equiv {\mathcal L}_1$ as in \eqref{eq:N=1_F_L}. However 
we notice that $\eta$ is of $m$-type for the ${\cal N}=2$ theory whereas it is of $q$-type for the ${\cal N}=1$ theory, which may puzzle the readers. 
In particular, the spontaneous breaking of a $q$-type symmetry necessarily gives rise to a pair of bosonic and fermionic vacua, leading to 
a vanishing Witten index $I_W^{\mathcal N=1}(A_{2}^{\rm def})=0$ (see more details in App.~\ref{App:N=1}).
This discrepancy can be resolved by the 't Hooft anomaly, which depends critically on the fermionic content:
\begin{itemize}
\item In $\mathcal N=1$, $\eta$ acts on a single Majorana fermion ($\nu=1$).
\item In $\mathcal N=2$, $\eta$ acts on a Dirac fermion, which is equivalent to two Majorana fermions ($\nu=2$).
\end{itemize}
This anomaly is classified by the spin cobordism group $\Omega_{2}^{\rm Spin}(B\mathbb Z_2)=\mathbb Z^\nu_8$ \cite{Kapustin:2014dxa, Chang:2022hud}. The $\mathcal N=2$ case ($\nu=2$) corresponds to an $m$-type object, resolving the paradox. This $m$-type nature is captured by the fermionic morphism $\eta \times \eta =\mathbb C^{0|1} I$.

More interestingly, an additional mixed 't Hooft anomaly between the $\eta$ and $(-1)^F$-lines enforces the solitonic states to have \emph{fractional fermion numbers}. The mixed anomaly is due to an ambiguity in resolving the 4-way junction between the two lines (The computation is summarized in App.~\ref{App:pentagon_anomaly}),
\begin{equation}
\begin{gathered}
\begin{tikzpicture}[scale=1]
            \draw [densely dashed, ultra thick, BurntOrange, -<-=.60] (1.2, 0) -- (1.2, 0.75);
            \draw [densely dashed, ultra thick, BurntOrange, -<-=.60] (1.2, .75) -- (1.2, 1.5);
 \node at (1.5, 1.4) {\footnotesize $\eta$};
 \node at (.5, 1.1) {\footnotesize $(-1)^F$};
 \draw [densely dashed, ultra thick, MidnightBlue, ->-=.6] (.2, 0.7) .. controls ( .7,.7) and (1.2,.7) .. (1.2, 0.55);
\draw [densely dashed, ultra thick, MidnightBlue, ->-=.6] (1.2, 0.9) .. controls (1.2, .75) and (1.8, .75) .. (2.3, 0.75);
\end{tikzpicture}
\end{gathered}
\quad = \quad (-1)\times\quad 
\begin{gathered}
\begin{tikzpicture}[scale=1]
            \draw [densely dashed, ultra thick, BurntOrange, -<-=.60] (1.2, 0) -- (1.2, 0.75);
            \draw [densely dashed, ultra thick, BurntOrange, -<-=.60] (1.2, .75) -- (1.2, 1.5);
 \node at (1.5, 1.4) {\footnotesize $\eta$};
 \node at (.5, 1.1) {\footnotesize $(-1)^F$};
 \draw [densely dashed, ultra thick, MidnightBlue, ->-=.6] (.2, 0.72) .. controls ( .7,.72) and (1.2,.72) .. (1.2, 0.87);
\draw [densely dashed, ultra thick, MidnightBlue, ->-=.6]  (1.2, 0.52) .. controls (1.2, .67) and (1.8, .67) .. (2.3, 0.67);
\end{tikzpicture}
\end{gathered}
\label{eq:anomaly}
\end{equation}
On the other hand, since the $\eta$-line is SSB, a solitonic state, e.g., in $\mathcal H_{I,\eta}$, is ``dressed" by an $\eta$-line. Therefore, the mixed anomaly forces that $(-1)^F$ must act projectively on the solitonic states:
\begin{align}
(-1)^{2F}|{\rm sol}\rangle\,&=\,
\begin{gathered}
\begin{tikzpicture}[scale=.9]
            \draw [thick, line] (0,-.1) -- (0,1.6);
            \draw [ultra thick, Cerulean, ->-=.6] (.2, 0) -- (1.2, 0);
            \draw [ultra thick, WildStrawberry, ->-=.7] (1.2, 0) -- (2.2, 0);
            \draw [densely dashed, ultra thick, BurntOrange] (1.2, 0) -- (1.2, 1.5);
            \draw [thick, line] (2.4,-.1) -- (2.6, .85) -- (2.4, 1.6);
 \filldraw[Cerulean] (1.2,0) circle (2pt);
 \node at (1.4, 1.45) {\footnotesize $\eta$};
 \node at (.6, 1.45) {\footnotesize $(-1)^F$};
 \draw [densely dashed, ultra thick, MidnightBlue, ->-=.6] (.2, 0.57) .. controls ( .7,0.57) and (1.2,0.57) .. (1.2, 0.42);
 \draw [densely dashed, ultra thick, MidnightBlue, ->-=.6] (1.2, 0.72) .. controls (1.2, 0.57) and (1.8, 0.57) .. (2.3, 0.57);
\draw [densely dashed, ultra thick, MidnightBlue, ->-=.6] (.2, 1.10) .. controls ( .7,1.10) and (1.2, 1.10) .. (1.2, .95);
 \draw [densely dashed, ultra thick, MidnightBlue, ->-=.6] (1.2, 1.25) .. controls (1.2, 1.1) and (1.8, 1.1) .. (2.3, 1.1);
\end{tikzpicture}
\end{gathered}
\,=\,-\,
\begin{gathered}
\begin{tikzpicture}[scale=.9]
            \draw [thick, line] (0,-.1) -- (0,1.6);
            \draw [ultra thick, Cerulean, ->-=.6] (.2, 0) -- (1.2, 0);
            \draw [ultra thick, WildStrawberry, ->-=.7] (1.2, 0) -- (2.2, 0);
            \draw [densely dashed, ultra thick, BurntOrange] (1.2, 0) -- (1.2, 1.5);
            \draw [thick, line] (2.4,-.1) -- (2.6, .85) -- (2.4, 1.6);
 \filldraw[Cerulean] (1.2,0) circle (2pt);
 \node at (1.4, 1.45) {\footnotesize $\eta$};
 \node at (.6, 1.45) {\footnotesize $(-1)^F$};
 \draw [densely dashed, ultra thick, MidnightBlue, ->-=.6] (.2, 0.57) .. controls ( .7,0.57) and (1.2,0.57) .. (1.2, 0.72);
 \draw [densely dashed, ultra thick, MidnightBlue, ->-=.6] (1.2, 0.42) .. controls (1.2, 0.57) and (1.8, 0.57) .. (2.3, 0.57);
\draw [densely dashed, ultra thick, MidnightBlue, ->-=.6] (.2, 1.10) .. controls ( .7,1.10) and (1.2, 1.10) .. (1.2, .95);
 \draw [densely dashed, ultra thick, MidnightBlue, ->-=.6] (1.2, 1.25) .. controls (1.2, 1.1) and (1.8, 1.1) .. (2.3, 1.1);
\end{tikzpicture}
\end{gathered}
\notag\\
&=\,-\,
\begin{gathered}
\begin{tikzpicture}[scale=.9]
            \draw [thick, line] (0,-.1) -- (0,1.6);
            \draw [ultra thick, Cerulean, ->-=.6] (.2, 0) -- (1.2, 0);
            \draw [ultra thick, WildStrawberry, ->-=.7] (1.2, 0) -- (2.2, 0);
            \draw [densely dashed, ultra thick, BurntOrange] (1.2, 0) -- (1.2, 1.5);
            \draw [thick, line] (2.5,-.1) -- (2.7, .85) -- (2.5, 1.6);
 \filldraw[Cerulean] (1.2,0) circle (2pt);
 \node at (1.4, 1.45) {\footnotesize $\eta$};
 \node at (.55, 1) {\footnotesize $(-1)^F$};
 \node at (2.15, .75) {\footnotesize $(-1)^F$};
\draw[densely dashed, ultra thick, MidnightBlue] (1.2, 1) arc (90:270:.2);
\draw[densely dashed, ultra thick, MidnightBlue] (1.2, 1.2) arc (90:-90:.4);
\end{tikzpicture}
\end{gathered}
=\,-\,|{\rm sol}\rangle
\end{align}
where the second-to-last equality was obtained by sequentially pinning together two $(-1)^F$'s on the left and the two on the right. We thus conclude that the (anti-)solitonic states must carry fractional fermion number,
\begin{align}
    F_{\text{(anti-)soliton}}=\pm\frac{1}{2}\,,
\end{align}
as a direct consequence of this mixed 't Hooft anomaly. We anticipate this result to extend to generic $A_{k+1}^{\rm def}$ models, with the anomaly persisting between the non-invertible line ${\mathcal L}_1$ and $(-1)^F$. We leave a comprehensive study of the generalization to arbitrary $k$
for future work.


This anomaly-based perspective elegantly explains an old result in \cite{Jackiw:1975fn,Goldstone:1981kk,Fendley:1991ve}. There, the fermion number for a soliton $s_{r,s}$ (connecting vacua $|r\rangle$ and $|s\rangle$) was computed from $\mathcal N=2$ supersymmetry as: $F_{r,s}=\frac{1}{2\pi}\mathfrak{Im}\log\frac{\partial^2_\Phi \mathcal W_{\mathcal N=2}(\Phi^{(r)})}{\partial^2_\Phi \mathcal W_{\mathcal N=2}(\Phi^{(s)})}$. Our result advocates that this fractionalization is fundamentally rooted in the hidden categorical symmetry of the theory.

\subsection{Quiver Representations and ADE classifications}
We now apply the superstrip algebra associated with $\tilde{\mathscr C}_{A_{k+1}}$ to understand the soliton spectra of the deformed $\mathcal N=2$ minimal models. As mentioned at the end of Sec.~\ref{Sec: catandalg}, we expect a \emph{connected diagram} encoding the IR spectrum. In the present case, the simplest connected quiver representation of the regular module is generated by the element ${\mathcal L}_1\in \tilde{\mathscr C}_{A_{k+1}}$. Following the fusion rule in \eqref{eq: su_2_N=2}, the action of $\mathcal L_1$ generates an $A$-type quiver connecting all $k+1$ vacua,
\begin{figure}[H]
\centering
    \begin{tikzpicture}[
    dot_style/.style={circle, fill, inner sep=1.5pt},
    blue_arc/.style={Cerulean, thick, -Stealth, bend left=25},
    red_arc/.style={WildStrawberry, thick, -Stealth, bend left=25},
    label_below/.style={below=2pt}, scale=0.80063
]

\node at (-1, 0) {$\mathcal R_{{\mathcal L}_1}$:};

    \node[dot_style, label={[label distance=5pt]below:$0$}] (n0) at (0, 0) {};
    \coordinate (n0pr) at (0.1, .125) {};
    \coordinate (n0pl) at (-0.9, .125) {};
    
    \coordinate (n0nr) at (0.1, -.125) {};
    \coordinate (n0nl) at (-0.9, -.125) {};

    \node[dot_style, label={[label distance=5pt]below:$1$}] (n1) at (1.5, 0) {};
    \coordinate (n1pr) at (1.6, .125) {};
    \coordinate (n1pl) at (1.4, .125) {};
    
    \coordinate (n1nr) at (1.6, -.125) {};
    \coordinate (n1nl) at (1.4, -.125) {};

    \node[dot_style, label={[label distance=5pt]below:$2$}] (n2) at (3.0, 0) {};
    \coordinate (n2pr) at (3.1, .125) {};
    \coordinate (n2pl) at (2.9, .125) {};

    \coordinate (n2nr) at (3.1, -.125) {};
    \coordinate (n2nl) at (2.9, -.125) {};

    \node[dot_style, label={[label distance=5pt]below:$3$}] (n3) at (4.5, 0) {};
    \coordinate (n3pr) at (4.6, .125) {};
    \coordinate (n3pl) at (4.4, .125) {};
        
    \coordinate (n3nr) at (4.6, -.125) {};
    \coordinate (n3nl) at (4.4, -.125) {};
    \node at (5.75, 0) {$\dots\dots$};

    \node[dot_style, label={[label distance=5pt]below:$k-1$}] (nk_minus_1) at (7.5, 0) {};
    \coordinate (nk1pr) at (7.6, .125) {};
    \coordinate (nk1pl) at (7.4, .125) {};

    \coordinate (nk1nr) at (7.6, -.125) {};
    \coordinate (nk1nl) at (7.4, -.125) {};
    
    \node[dot_style, label={[label distance=5pt]below:$k$}] (nk) at (9, 0) {};
    \coordinate (nkpr) at (9.1, .125) {};
    \coordinate (nkpl) at (8.9, .125) {};

    \coordinate (nknr) at (9.1, -.125) {};
    \coordinate (nknl) at (8.9, -.125) {};

    \draw[blue_arc] (n0pr) to (n1pl);
    \draw[red_arc] (n1nl) to (n0nr);
    \draw[red_arc] (n1pr) to (n2pl);
    \draw[blue_arc] (n2nl) to (n1nr);
    \draw[blue_arc] (n2pr) to (n3pl);
    \draw[red_arc] (n3nl) to (n2nr);
    \draw[red_arc] (nk1pr) to (nkpl);
    \draw[blue_arc] (nknl) to (nk1nr);
\end{tikzpicture}
\end{figure}
\noindent Here, the blue/red lines represent the bosonic/fermionic junctions arising from the fusion of boundaries with the ${\mathcal L}_1$-line. The quiver encodes the morphisms between Hilbert space sub-sectors defined by vacua $|r\rangle$ and $|r+1\rangle$ at $\sigma=\pm \infty$, and, consequently the \emph{boson/fermion degeneracies} within those sectors:
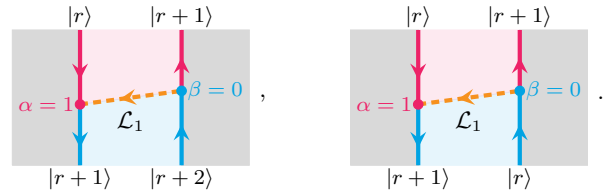
\begin{figure}[H] 
    \centering 
    \begin{subfigure}{\columnwidth}
        \centering 
    \begin{minipage}{0.48\linewidth}
    \begin{tikzpicture}[scale=.9]
        \filldraw[gray,opacity=0.3] (0,0)--(1,0)--(1,2)--(0,2)--(0,0);
        \filldraw[gray,opacity=0.3] (2.5,0)--(3.5,0)--(3.5,2)--(2.5,2)--(2.5,0);
        \filldraw[Cerulean,opacity=0.1] (1,0)--(1,0.9)--(2.5,1.1)--(2.5,0)--(1,0);
        \filldraw[WildStrawberry,opacity=0.1] (1,0.9)--(2.5,1.1)--(2.5,2)--(1,2)--(1,0.9);
        \draw[ultra thick,Cerulean,-<-=.55] (1,0)--(1,.9);
        \draw[ultra thick,WildStrawberry,-<-=.55] (1,.9)--(1,2);
        \draw[ultra thick,Cerulean,->-=.55] (2.5,0)--(2.5,1.1);
        \draw[ultra thick,WildStrawberry,->-=.55] (2.5,1.1)--(2.5,2);
        \draw[densely dashed,ultra thick,BurntOrange,-<-=.55] (1,0.9)--(2.5,1.1);
        \node at (0.5,.9) {\footnotesize$\color{WildStrawberry}\alpha=1$};
        \node at (3,1.1) {\footnotesize$\color{Cerulean}\beta=0$};
        \filldraw[WildStrawberry] (1,0.9) circle (2pt);
        \filldraw[Cerulean] (2.5,1.1) circle (2pt);
        \node at (1., -.2) {\footnotesize$|r+1\rangle$};
        \node at (1.,2.2) {\footnotesize $|r\rangle$};
        \node at (2.5,-.2) {\footnotesize$|r+2\rangle$};
        \node at (2.5,2.2) {\footnotesize $|r+1\rangle$};
        \node at (1.75,0.65) {${\mathcal L}_1$};
        \node at (3.7, 1) {,};
    \end{tikzpicture}
    \end{minipage}
        \hfill 
        \begin{minipage}{0.48\linewidth}
 \begin{tikzpicture}[scale=.9]
        \filldraw[gray,opacity=0.3] (0,0)--(1,0)--(1,2)--(0,2)--(0,0);
        \filldraw[gray,opacity=0.3] (2.5,0)--(3.5,0)--(3.5,2)--(2.5,2)--(2.5,0);
        \filldraw[Cerulean,opacity=0.1] (1,0)--(1,0.9)--(2.5,1.1)--(2.5,0)--(1,0);
        \filldraw[WildStrawberry,opacity=0.1] (1,0.9)--(2.5,1.1)--(2.5,2)--(1,2)--(1,0.9);
        \draw[ultra thick,Cerulean,-<-=.55] (1,0)--(1,.9);
        \draw[ultra thick,WildStrawberry,-<-=.55] (1,.9)--(1,2);
        \draw[ultra thick,Cerulean,->-=.55] (2.5,0)--(2.5,1.1);
        \draw[ultra thick,WildStrawberry,->-=.55] (2.5,1.1)--(2.5,2);
        \draw[densely dashed,ultra thick,BurntOrange,-<-=.55] (1,0.9)--(2.5,1.1);
        \node at (0.5,.9) {\footnotesize$\color{WildStrawberry}\alpha=1$};
        \node at (3,1.1) {\footnotesize$\color{Cerulean}\beta=0$};
        \filldraw[WildStrawberry] (1,0.9) circle (2pt);
        \filldraw[Cerulean] (2.5,1.1) circle (2pt);
        \node at (1., -.2) {\footnotesize$|r+1\rangle$};
        \node at (1.,2.2) {\footnotesize $|r\rangle$};
        \node at (2.5,-.2) {\footnotesize$|r\rangle$};
        \node at (2.5,2.2) {\footnotesize $|r+1\rangle$};
        \node at (1.75,0.65) {${\mathcal L}_1$};
        \node at (3.7, 1) {.};
    \end{tikzpicture} 
        \end{minipage}
    \end{subfigure}
    \caption{Degeneracies between fermionic/bosonic states in soliton/soliton
     sectors $\mathcal H_{r,r+ 1}\leftrightarrows\mathcal H_{r+1,r+2}$, and soliton/anti-soliton ones $\mathcal H_{r,r+1}\leftrightarrows\mathcal H_{r+1,r}$ for odd $r$.}
    \label{fig:degeneracy}
\end{figure}
\noindent
From this representation, we conclude the following properties regarding the spectra of the deformed $A_{k+1}^{\rm def}$ models:
\begin{itemize}
\item The $k+1$ vacua contain no local particle excitations; all gapped states are solitons connecting adjacent vacua. 
\item In each (anti-)soliton sector $H_{r,r+1}$ ($\mathcal H_{r+1,r}$), there is at least a single bosonic or fermionic soliton state, denoted by $b_{r,r+1}$ or $f_{r,r+1}$ ($\bar b_{r+1,r}$ or $\bar f_{r+1,r}$), taking opposite fractional fermion number $F_{b,f}=\mp\frac{1}{2}$.
\item These states must have the same mass, 
{
\begin{align}
\cdots=m_{\bar f_{r, r-1}}=m_{b_{r-1,r}}=m_{f_{r,r+1}}=m_{\bar b_{r+1,r}}=\cdots\notag
\end{align}
}
\end{itemize}
In accordance with the unbroken supersymmetry, any state (e.g. $f_{r,r+1}$) must have a super-partner ($b_{r,r+1}$ ) within the same Hilbert space sector $\mathcal H_{r,r+1}$. They together form an irreducible $\mathcal N=2$ massive short multiplet $s_{r,r+1} =(b_{r,r+1},f_{r,r+1})$. Therefore, the non-invertible symmetries provide constraints ``orthogonal" to supersymmetry: SUSY dictates the bosonic/fermionic pairing within each sector, while the non-invertible symmetries organize the $2k$ (anti-)solitonic SUSY multiplets into a multiplet of the superstrip algebra. This result, combined with supersymmetry, is in remarkable agreement with integrability literature \cite{Bernard:1990ti,Fendley:1991ve} (see also the beginning in Sec. \ref{sec:ADE}).\\

\paragraph{$D/E$-type least-relevantly-deformed $\mathcal N=2$ minimal models}
We now turn to the deformed $D/E$-type $\mathcal N=2$ minimal models. At the conformal point, these $D/E$-type models are known generalized orbifolds of specific $A$-type theories: the $D_{k+2}$ models from $A_{2k+1}$, and the $E_{6,7,8}$ models from $A_{11,17,29}$ respectively. This relationship is realized in each case by gauging non-invertible symmetries corresponding to a Frobenius algebra, $\mathscr A_{D/E}$ \cite{Fuchs:2002cm, Carqueville:2012dk, Diatlyk:2023fwf}. Crucially, the algebra $\mathscr A_{D/E}$ survives the least relevant deformation and is preserved within the symmetry category of the deformed $A$-type model, $\tilde{\mathscr C}_A$ associated with $A_{2k+1}^{\rm def}$. Therefore, the $D/E$-type gapped theory can be realized by gauging the symmetries defined by this algebra $\mathscr A_{D/E}$ within the deformed $A$-type theory—an operation that remains valid throughout the entire RG flow.

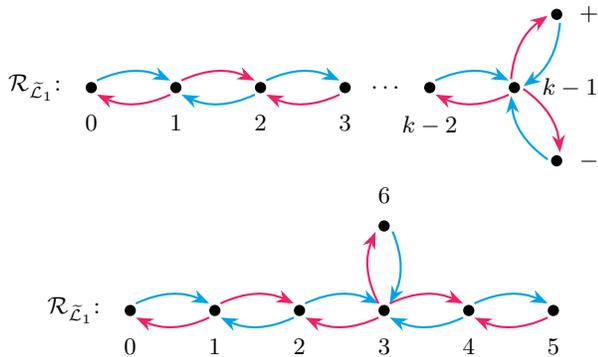
\begin{figure}
\centering
    \begin{tikzpicture}[
    dot_style/.style={circle, fill, inner sep=1.5pt},
    blue_arc/.style={Cerulean, thick, -Stealth, bend left=25},
    red_arc/.style={WildStrawberry, thick, -Stealth, bend left=25},
    label_below/.style={below=2pt}, scale=0.75
]

\node at (-1, 0) {$\mathcal R_{\widetilde{\mathcal L}_1}$:};

    \node[dot_style, label={[label distance=5pt]below:$0$}] (n0) at (0, 0) {};
    \coordinate (n0pr) at (0.1, .125) {};
    \coordinate (n0pl) at (-0.9, .125) {};
    
    \coordinate (n0nr) at (0.1, -.125) {};
    \coordinate (n0nl) at (-0.9, -.125) {};

    \node[dot_style, label={[label distance=5pt]below:$1$}] (n1) at (1.5, 0) {};
    \coordinate (n1pr) at (1.6, .125) {};
    \coordinate (n1pl) at (1.4, .125) {};
    
    \coordinate (n1nr) at (1.6, -.125) {};
    \coordinate (n1nl) at (1.4, -.125) {};
    \node[dot_style, label={[label distance=5pt]below:$2$}] (n2) at (3.0, 0) {};
    \coordinate (n2pr) at (3.1, .125) {};
    \coordinate (n2pl) at (2.9, .125) {};

    \coordinate (n2nr) at (3.1, -.125) {};
    \coordinate (n2nl) at (2.9, -.125) {};
    \node[dot_style, label={[label distance=5pt]below:$3$}] (n3) at (4.5, 0) {};
    \coordinate (n3pr) at (4.6, .125) {};
    \coordinate (n3pl) at (4.4, .125) {};
        
    \coordinate (n3nr) at (4.6, -.125) {};
    \coordinate (n3nl) at (4.4, -.125) {};
    \node at (5.25, 0) {$\dots$};

    \node[dot_style, label={[label distance=5pt]below:$k-2$}] (k2) at (6, 0) {};
    \coordinate (nk2pr) at (6.1, .125) {};
    \coordinate (nk2pl) at (-6.9, .125) {};
    
    \coordinate (nk2nr) at (6.1, -.125) {};
    \coordinate (nk2nl) at (-6.9, -.125) {};

\node[dot_style, label={[label distance=5pt]right:$k-1$}] (nk1) at (7.5, 0) {};
    
    \coordinate (nk1pl) at (7.4, .125) {};
    \coordinate (nk1nl) at (7.4, -.125) {};
    
    \coordinate (nk1pr) at (7.442, .149) {};
    \coordinate (nk1nr) at (7.658, 0.024) {};

    \coordinate (nk1pr2) at (7.658, -0.024) {};
    \coordinate (nk1nr2) at (7.442, -.149) {};

    \node[dot_style, label={[label distance=3pt]right:$+$}] (np) at (8.25, 1.299) {};
    \coordinate (nppl) at (8.092, 1.275) {};
    \coordinate (npnl) at (8.308, 1.150) {};
    \node[dot_style, label={[label distance=3pt]right:$-$}] (nn) at (8.25, -1.299) {};
    \coordinate (nnpl) at (8.308, -1.150) {};
        
    \coordinate (nnnl) at (8.092, -1.275) {};
   
    \draw[blue_arc] (n0pr) to (n1pl);
    \draw[red_arc] (n1nl) to (n0nr);
    \draw[red_arc] (n1pr) to (n2pl);
    \draw[blue_arc] (n2nl) to (n1nr);
    \draw[blue_arc] (n2pr) to (n3pl);
    \draw[red_arc] (n3nl) to (n2nr);
    \draw[blue_arc] (nk2pr) to (nk1pl);
    \draw[red_arc] (nk1nl) to (nk2nr);
    \draw[red_arc] (nk1pr) to (nppl);
    \draw[blue_arc] (npnl) to (nk1nr);
    \draw[red_arc] (nk1pr2) to (nnpl);
    \draw[blue_arc]  (nnnl) to (nk1nr2);
\end{tikzpicture}
\begin{tikzpicture}[
    dot_style/.style={circle, fill, inner sep=1.5pt},
    blue_arc/.style={Cerulean, thick, -Stealth, bend left=25},
    red_arc/.style={WildStrawberry, thick, -Stealth, bend left=25},
    label_below/.style={below=2pt}, scale=0.75
]

\node at (-1, 0) {$\mathcal R_{\widetilde{\mathcal L}_1}$:};

    \node[dot_style, label={[label distance=5pt]below:$0$}] (n0) at (0, 0) {};
    \coordinate (n0pr) at (0.1, .125) {};
    \coordinate (n0pl) at (-0.9, .125) {};
    
    \coordinate (n0nr) at (0.1, -.125) {};
    \coordinate (n0nl) at (-0.9, -.125) {};

    \node[dot_style, label={[label distance=5pt]below:$1$}] (n1) at (1.5, 0) {};
    \coordinate (n1pr) at (1.6, .125) {};
    \coordinate (n1pl) at (1.4, .125) {};
    
    \coordinate (n1nr) at (1.6, -.125) {};
    \coordinate (n1nl) at (1.4, -.125) {};
    \node[dot_style, label={[label distance=5pt]below:$2$}] (n2) at (3.0, 0) {};
    \coordinate (n2pr) at (3.1, .125) {};
    \coordinate (n2pl) at (2.9, .125) {};

    \coordinate (n2nr) at (3.1, -.125) {};
    \coordinate (n2nl) at (2.9, -.125) {};
    \node[dot_style, label={[label distance=5pt]below:$3$}] (n3) at (4.5, 0) {};
    \coordinate (n3pr) at (4.6, .125) {};
    \coordinate (n3pl) at (4.4, .125) {};
        
    \coordinate (n3nr) at (4.6, -.125) {};
    \coordinate (n3nl) at (4.4, -.125) {};

    \node[dot_style, label={[label distance=5pt]below:$4$}] (n4) at (6, 0) {};
    \coordinate (n4pr) at (6.1, .125) {};
    \coordinate (n4pl) at (5.9, .125) {};
        
    \coordinate (n4nr) at (6.1, -.125) {};
    \coordinate (n4nl) at (5.9, -.125) {};

    \node[dot_style, label={[label distance=5pt]below:$5$}] (n5) at (7.5, 0) {};
    \coordinate (n5pr) at (7.6, .125) {};
    \coordinate (n5pl) at (7.4, .125) {};
        
    \coordinate (n5nr) at (7.6, -.125) {};
    \coordinate (n5nl) at (7.4, -.125) {};

    \node[dot_style, label={[label distance=3pt]above:$6$}] (n6) at (4.5, 1.5) {};
    \coordinate (n6pl) at (4.35, 1.4) {};
        
    \coordinate (n6nl) at (4.65, 1.4) {};

    \draw[blue_arc] (n0pr) to (n1pl);
    \draw[red_arc] (n1nl) to (n0nr);
    \draw[red_arc] (n1pr) to (n2pl);
    \draw[blue_arc] (n2nl) to (n1nr);
    \draw[blue_arc] (n2pr) to (n3pl);
    \draw[red_arc] (n3nl) to (n2nr);
    \draw[red_arc] (n3pr) to (n4pl);
    \draw[blue_arc] (n4nl) to (n3nr);
    
    \draw[red_arc] (n3pl) to (n6pl);
    \draw[blue_arc] (n6nl) to (n3pr);

    \draw[blue_arc] (n4pr) to (n5pl);
    \draw[red_arc] (n5nl) to (n4nr);
\end{tikzpicture}
\caption{The $\mathcal R_{\widetilde{\mathcal L}_1}$ superstrip algebra representation for the $D^{\rm def}_{k+2}$ and $E^{\rm def}_7$ models. }
\label{fig:D/E_quivers}
\end{figure}

On the categorical level, the vacua states and their associated spontaneously broken symmetries in the $D^{\rm def}_{k+2}/E^{\rm def}_{6,7,8}$ models are uniquely determined by the left $\mathscr A_{D/E}$-module of $\tilde{\mathscr C}_A$,
\begin{align}
    \tilde{\mathscr C}_{D/E}=\tilde{\mathscr C}_A/\mathscr A_{D/E}\,.
\end{align}
All resulting $\tilde{\mathscr C}_{D/E}$ categories are generated by a simple object, $\widetilde{\mathcal L}_1\equiv\mathcal L_1\otimes\mathscr A_{D/E}$, which is itself constructed from the generator $\mathcal L_1\in \tilde{\mathscr C}_A$ and the algebra $\mathscr A_{D/E}$ (computation details are collected in App.~\ref{App: orbifold}). The non-invertible symmetry $\widetilde{\mathcal L}_1$ directly corresponds to a $D/E$-type superstrip algebra representation, which in turn encodes the full vacua structure and degenerate (anti)-soliton spectra. In Fig.~\ref{fig:D/E_quivers},
The rep-$\mathcal R_{\widetilde{\mathcal L}_1}$ reveals that the deformed minimal models possess a vacuum structure precisely matching the $D/E$-type Dynkin diagrams \cite{Fendley:1991ve}. The fundamental excitations are mass-degenerate (anti-)solitons connecting adjacent vacua, each composed of a bosonic-fermionic pair with fractional fermion numbers $F_{b,f}=\mp1/2$.

\section{Discussions}
In this letter, we introduced the superstrip algebra as a general framework for 2D gapped fermionic systems possessing fermionic categorical symmetries. We demonstrated its power by analyzing the deformed $\mathcal N=2$ minimal models. We showed that a non-invertible super-fusion category $\tilde{\mathscr C}_{A/D/E}$, preserved only by the least relevant deformation, is the fundamental origin of the rigid $ADE$ solitonic spectrum. This symmetry-based approach, independent of supersymmetry or integrability, dictates the fermionic/bosonic mass degeneracies and enforces the $\frac{1}{2}$-fractional fermion numbers via a mixed 't Hooft anomaly. 

Our framework opens numerous new avenues:
\begin{itemize}
\item 
First, it naturally extends to other 2D fermionic deformed CFTs, e.g. the $\mathcal N=1$ minimal models and $\mathcal N=2$ Kazama Suzuki models \cite{Ahn:1990gn, toappear, Fendley:1993pi} under their least relevant deformations. These systems, being both integrable and supersymmetric, provide an ideal testing ground to further validate the robustness of our approach and to uncover the hidden categorical symmetries governing their known solitonic spectra.

\item 
Another vital application is investigating how fermionic non-invertible symmetries constrain integrable S-matrices. It was recently realized \cite{Copetti:2024rqj} that crossing symmetry equations in bosonic systems must be modified with respect to non-invertible symmetry data. Generalizing it to the fermionic realm is imperative. Our framework provides the necessary ingredients (e.g., superfusion rules and $F$-symbols) to derive these modified crossing relations, establishing a new bootstrap program for fermionic integrable models.

\item 
Significantly, our framework applies to systems lacking both supersymmetry and integrability. A prime candidate is 2D massless adjoint QCDs. While recent breakthroughs \cite{Komargodski:2020mxz, Cordova:2024goh, Cordova:2024nux} have utilized bosonized duals to constrain deconfinement and spectra of the IR physics, the native symmetries are intrinsically fermionic. Applying our superstrip algebra to these fermionic categories promises to refine these constraints and shed new light on the non-perturbative dynamics.

\item 
Finally, we highlight the connection to 4D physics via superconformal vortex strings \cite{Tong:2006pa}, which relates the Coulomb branch operators in 4D $\mathcal N=2$ Argyres-Douglas (AD) theories to chiral ring operators in $\mathcal N=2$ minimal models. Based on 
the above 2D/4D correspondence, one can find special loci in the minimal chamber of the AD theory where the BPS spectrum consists of $k$ monopoles of equal mass, precisely matching the BPS solitons of the corresponding minimal model with the least relevant deformation \cite{Shapere:1999xr}. In fact, each soliton can be viewed as a monopole confined in the core of the vortex string \cite{Tong:2006pa}. The result of the present work then implies a striking corollary for four dimensions: we suspect that 
a (hidden) non-invertible symmetry should exist in four dimensions to underlie such 4D BPS degeneracies at the special loci. 


\end{itemize}

%
%
%
%
%
%

\section*{Acknowledgment}
We thank Mahesh Balasubramanian, Matthew Buican, Zohar Komargodski, Piljin Yi, Yunqin Zheng for useful discussions. J.C. is supported by the Fundamental Research Funds for the Central Universities (No.20720230010) of China, Fujian Provincial Natural Science Foundation of China (No.2025J01004), and the National Natural Science Foundation of China (Grants No.12247103). Z.D. is supported by a STFC Consolidated Grant, ST$\backslash$T000686$\backslash$1 ``Amplitudes, strings \& duality" and also the Swiss National Science Foundation (SNSF) through a Postdoctoral Fellowship (TMPFP2\_234009). S.L. is supported by KIAS Grant PG056502. QJ is supported by National Research Foundation of Korea (NRF) Grant No. RS-2024-00405629 and Jang Young-Sil Fellow Program at the Korea Advanced Institute of Science and Technology.

\bibliographystyle{apsrev4-2}

\bibliography{main.bib}




\section*{Supplemental Material}

\begin{appendices}

\section{Fermionization, Condensations and Vacua}
\label{app:Fermionization}

In this appendix, we will review how to perform fermion condensation on a 2D bosonic theory $\mathfrak{T}_b$, which is equivalent to fermionization, and extract the corresponding categorical symmetry in the resulting fermionic theory $\mathfrak{T}_f$. 

Suppose there exists a non-anomalous $\mathbb{Z}_2$ symmetry in $\mathfrak{T}_b$, represented by a set of TDLs $\{\mathcal{L}_0,\mathcal{L}_1\}$, where $\mathcal{L}_0$ is the identity and $\mathcal{L}_1$ is the $\mathbb{Z}_2$ generator. We place the theory $\mathfrak{T}_b$ on a torus and introduce the partition function $Z_b[a_t,a_s]$ as
    \begin{equation}
        Z_b[a_t,a_s] = \textrm{Tr}_{\mathcal{H}^b_{a_s}} \mathcal{L}_1^{a_t} e^{-\beta H^b}\,,
    \end{equation}
where $a_t,a_s=0,1$ denote the temporal and spatial $\mathbb{Z}_2$ holonomies, respectively, introduced by inserting the TDL $\mathcal{L}_1$ along spatial and temporal circles. The insertion of $\mathcal{L}_1$ along temporal circle modifies the theory $\mathfrak{T}_b$ and change the Hilbert space $\mathcal{H}^b_0$ to a twist Hilbert space $\mathcal{H}^b_1$. On the other hand, the $\mathcal{L}_1$ along spatial circle will act on the Hilbert space and is represented by inserting $\mathcal{L}_1$ in the trace. After fermionization, the fermionic partition function $Z_f[s_1,s_2]$ of $\mathfrak{T}_f$ is 
    \begin{equation}
        Z_f[s_t,s_s] = \frac{1}{2} \sum_{a_t,a_s=0,1} (-1)^{s_t a_s + s_s a_t + a_t a_s} Z_b[a_t,a_s]\,,
    \end{equation}
where $s_t,s_s=0,1$ denote the temporal and spatial spin structures, which are shifted by inserting the fermion parity $(-1)^F$ in the same way. Conventionally, $s_s = 0$ is called the Neveu-Schwarz sector and $s_s = 1$ is known as the Ramond sector. Hence we have
    \begin{equation}
        Z_f[s_t,s_s] = \textrm{Tr}_{\mathcal{H}^f_{s_s}} (-1)^{s_t F} e^{-\beta H^f}\,.
    \end{equation}
In particular, the Witten index is given by the partition function $Z_f(s_s=1,s_t=1)$.

The TDLs in $\mathfrak{T}_f$ other than $(-1)^F$ can be obtained as follows. Notice that all TDLs in $\mathfrak{T}_b$ are classified according to their $\mathbb{Z}_2$-orbits, and we label them as $\mathcal{L}_i$ with $i=0,\cdots,2m+q-1$ and $m,q\in \mathbb{Z}_{\geq 0}$. Among them, the pair $\{\mathcal{L}_{2k-2},\mathcal{L}_{2k-1}\}$ with $k=1,\cdots,m$ form a doublet under $\mathbb{Z}_2$-action
    \begin{equation}
        \mathcal{L}_1 \cdot \mathcal{L}_{2k}=\mathcal{L}_{2k+1}\,,\quad \mathcal{L}_1 \cdot \mathcal{L}_{2k+1} = \mathcal{L}_{2k}\,,
    \end{equation}
while $\mathcal{L}_{2m+l-1}$ with $l=1,\cdots,q$ are fixed point
    \begin{equation}
        \mathcal{L}_1 \cdot \mathcal{L}_{2m+l-1}= \mathcal{L}_{2m+l-1}\,.
    \end{equation}
Upon fermionization, the super fusion category of $\mathfrak{T}_f$ consists of the TDLs $\widetilde{\mathcal{L}}_{a}$ with $a=0,\cdots,m+q-1$, where $\widetilde{\mathcal{L}}_{k-1}$ with $k=1,\cdots,m$ are $m$-type TDLs representing the orbits $\{\mathcal{L}_{2k-2},\mathcal{L}_{2k-1}\}$, while $\widetilde{\mathcal{L}}_{m-1+l}$ with $l=1,\cdots,q$ are $q$-type TDLs corresponding to the fixed points $\mathcal{L}_{2m+l-1}$. For simplicity, we may also omit the tilde if no confusion occurs.

We will examine a special case where $\mathfrak{T}_b$ is gapped and the symmetry category $\mathscr{C}_b$ generated by all TDLs $\mathcal{L}_i$ is spontaneously broken, resulting in $2m+q$ vacua. The bosonic partition function $Z_b[0,0]=2m+q$ simply counts the total number of vacua. Moreover, when the temporal holonomy is turned on $(a_t=1)$, we have $Z_b[1,0]=q$ since only the vacua associated to $\mathcal{L}_{2m-1+l}$ remain invariant under the $\mathbb{Z}_2$ action and thus contribute the trace. Applying modular $S/T$-transformation on the torus, we obtain the partition functions in other sectors, and we summarize
    \begin{equation}
        Z_b[0,0]=2m+q\,, \quad Z_b[1,0]=Z_b[0,1]=Z_b[1,1]=q\,.
    \end{equation}
Applying fermionization, we obtain the partition function of $\mathfrak{T}_f$
    \begin{equation}
        Z_f[s_t,s_s] = m+(-1)^{s_t s_s} q\,.
    \end{equation}
Here the total number of vacua is $m+q$, implying the entire superfusion category $\mathscr{C}_f$, except the fermion parity $(-1)^F$ that is factorized out, is spontaneously broken, and the vacua are in one-to-one correspondent to the TDLs $\widetilde{\mathcal{L}}_a$ labeled by $a=0,\cdots,m+q-1$. In the Ramond sector $(s_s=1)$, the sign factor $(-1)^{s_t s_s}$ indicates the vacua associated to $m$-type and $q$-type TDLs are respectively bosonic and fermionic. 

Let us consider the example of $\mathcal{N}=2$ $A_{k+1}$ minimal model. The bosonization of the model is $\frac{{\rm SU}(2)_k\times {\rm U}(1)_2}{{\rm U}(1)_{k+2}}$. It contains $2(k+1)(k+2)$ bosonic primaries $\phi_{(a,c)}$ along with their corresponding Verlinde lines $\mathcal L_{(a,c)}$ furnishing a fusion category denoted by $\mathscr C_b$, where $a=0,1,\dots,k$ and $c\in\mathbb Z_{2k+4}$. Here we follow the notations in \cite{Gray:2008je, Cordova:2023qei}. The modular $S$-matrix is spelled as
\begin{align}
     S_{(a,c),(a',c')}=&\frac{1}{k+2}\sin\left(\frac{\pi(a+1)(a'+1)}{k+2}\right)\notag\\
                       &\cdot e^{i\pi\frac{cc'}{k+2}}e^{-i\pi\frac{[a+c][a'+c']}{2}}\,,
\end{align}
where $[x]\equiv x\ ({\rm mod}\ 2)$. The fermionic supersymmetric model $A_{k+1}$ can be obtained again via fermionizing the line $\mathcal L_{(k,k+2)}$.

In $A_{k+1}$, there are $k+1$ relevant superconformal primary operator in NS-sector, denoted by $\Phi_{(l,l)}$, for $l=0,1,\dots,k$. Notice that they are $\mathcal N=2$ short multiplets, and have (holomorphic) conformal weight equal to half of their ${\rm U}(1)_R$-charge $q_{(l,l)}$,
\begin{align}
    h_{(l,l)}=\frac{q_{(l,l)}}{2}=\frac{l}{2k+4}.
\end{align}
In the bosonic sub-algebra, $\Phi_{(l,l)}$ is decomposed to 
\begin{align}
    \Phi_{(l,l)}\equiv\left(\phi_{(l,l)},\,\phi_{(k-l,l+k+2)}\right)\,.
\end{align}
Therefore $\Phi_{(k,k)}\equiv\left(\phi_{(k,k)},\,\phi_{(0,2k+2)}\right)$ is the least relevant operator. The deformation with respect to it, after integrating over the superspace, only matters with the top component $\phi_{(0,2k+2)}$. To determine the preserved TDLs under $\phi_{(0,2k+2)}$, our strategy is to first compute the preserved fusion category $\tilde{\mathscr C}_b$ in the deformed bosonic theory, denoted by $\mathcal B\left[A_{k+1}^{\rm def}\right]$. By a further fermionic anyon condensation, one can obtain the preserved superfusion category $\tilde{\mathscr C}_{A_{k+1}}$.

Therefore, for the operator $\Phi_{(k,k)}^{\rm top}=\phi_{(0,2k+2)}$, the preserved TDLs need to satisfy following condition \cite{Chang:2018iay}:
\begin{equation}
\begin{gathered}
\begin{tikzpicture}[scale=.6]
\draw [line,ultra thick,dashed,MidnightBlue] (-.5,-0.75) circle (1.125) ;
\draw (0,-2.5) node[scale=.6] {$\mathcal{L}_{(a,c)}$};
\draw (0,1.2) node {};
\draw (-0.5,-0.75)\dotsols{[scale=.6, outer sep=0pt]below=0pt}{$\phi_{(0,\,2k+2)}$};
\draw [line,ultra thick,DarkBlue,-<-=1.0](-.47, 0.375) -- (-.46, 0.375);
\draw [line,ultra thick,DarkBlue,-<-=1.0](-.48, -1.875)--(-.49, -1.875);
\end{tikzpicture}
\end{gathered}\quad = \quad
\begin{gathered}
\begin{tikzpicture}[scale=.6]
\draw [line,ultra thick,dashed,MidnightBlue] (-.5,-0.75) circle (1.125) ;
\draw (0,-2.5) node[scale=.6] {$\mathcal{L}_{(a,c)}$};
\draw (0,1.2) node {};
\draw (1.7,-0.75)\dotsols{[scale=.6, outer sep=0pt]below=0pt}{$\phi_{(0,\,2k+2)}$};
\draw [line,ultra thick,DarkBlue,-<-=1.0](-.47, 0.375) -- (-.46, 0.375);
\draw [line,ultra thick,DarkBlue,-<-=1.0](-.48, -1.875)--(-.49, -1.875);
\end{tikzpicture}
\end{gathered}
=
\left\langle\mathcal L_{(a,\,c)}\right\rangle|\phi_{(0,\,2k+2)}\rangle
\,,
\end{equation}
i.e. the charge of $\mathcal L_{(a,\,c)}$ on $\phi_{(0,\,{2k+2})}$ equals its own quantum dimension,
\begin{align}
    \left\langle\mathcal L_{(a,\,c)}\right\rangle=\frac{S_{(a,c),(0,0)}}{S_{(0,0),(0,0)}}=Q_{(a,c)}\left(\phi_{(0,2k+2)}\right)=\frac{S_{(a,c),(0,2k+2)}}{S_{(0,0),(0,2k+2)}}\,.
\end{align}
One then finds, in $\mathcal B[A_{k+1}^{\rm def}]$, the preserved fusion category 
\begin{align}
\tilde{\mathscr  C}_b=\{\,\mathcal L_{(a,\,0)},\ \mathcal L_{(a,\,k+2)}|a=0,1,\dots,k\}\,.
\end{align}
Their fusion rule can be determined by the general ones in the bosonized $\frac{{\rm SU}(2)_k\times {\rm U}(1)_2}{{\rm U}(1)_{k+2}}$,
{\small
\begin{align}
   &\mathcal L_{(a,\,c)} \cdot \mathcal L_{(a',\,c')} = \sum_{(\alpha, \gamma)} N_{(a,c),(a',c')}^{(\alpha, \gamma)} \mathcal L_{(\alpha,\, \gamma)}\,,\\\notag
   &N_{(a,c),(a',c')}^{(\alpha, \gamma)} =
\begin{cases}
(N_{\text{SU(2)}_k})_{a,a'}^{\alpha} (N_{\text{U(1)}_{k+2}})_{c,c'}^{\gamma}\\ 
\qquad\qquad\qquad\qquad\quad\ \text{if } [a+c]\cdot[a'+c']=0 \\\\
(N_{\text{SU(2)}_k})_{a,a'}^{k-\alpha} (N_{\text{U(1)}_{k+2}})_{c,c'}^{\gamma+k+2}\\ 
\qquad\qquad\qquad\qquad\quad\  \text{if } [a+c]\cdot[a'+c']=1
\end{cases}
\end{align}
}
\!\!\!\!\! where $N_{\text{SU(2)}_k}$ and $N_{\text{U(1)}_{k+2}}$ are the fusion matrices of Kac-Moody algebra $\text{SU(2)}_k$ and $\text{U(1)}_{k+2}$ respectively. To further perform the fermionic condensation with $\mathcal{L}_{(k,k+2)}$, it is necessary to notice that
\begin{equation}
    \mathcal L_{(a,c)}\cdot\mathcal L_{(k,k+2)}=\mathcal L_{(k-a,c+k+2)}\,.
    \label{eq:Z2_on_C}
\end{equation}
Since $\mathcal L_{(k,\,k+2)}$ acts transitively on $\tilde{\mathcal C}_b$, we denote $\mathcal L_a\equiv \mathcal L_{(a,0)}$ as the representative of the orbit $\{\mathcal L_{(a,\, 0)},\mathcal L_{(k-a,\, k+2)}\}$. They furnish the condensed superfusion category $\tilde{\mathscr C}_{A_{k+1}}=\{\,\mathcal L_a\,|a=0,1,2,\dots k\}$ in $A_{k+1}^{\rm def}$ with fusion rule:
\begin{equation}
    \mathcal L_a\cdot \mathcal L_{a'}=
    \begin{cases}
        \mathbb C^{1|0}\sum_{\alpha=|a-a'|\,,\,{\rm step}\ 2}^{\min(a+a',\, 2k-a-a')}  \mathcal L_\alpha\,,\quad {\rm if}\ \ a\cdot a'\ {\rm even}\,, \\[5pt]
        \mathbb C^{0|1}\sum_{\alpha=|a-a'|\,,\,{\rm step}\ 2}^{\min(a+a',\, 2k-a-a')}  \mathcal L_\alpha\,,\quad {\rm if}\ \ a\cdot a'\ {\rm odd}\,,
    \end{cases}
    \label{eq: su_2_N=2}
\end{equation}
where the fermionic junction is determined when the fusion of two objects $\mathcal L_a$ and $\mathcal L_{a'}$ is not in the picked representative set $\{\mathcal L_{(a,0)}\,|\,a=0,1,\dots,k\}$, e.g.
\begin{align}
    \mathcal L_{(1,0)}\cdot \mathcal L_{(1,0)}&=\mathbb C^{1|0}\,\mathcal L_{(k,k+2)}+\mathbb C^{1|0}\,\mathcal L_{(k-2,k+2)}\notag\\
    &\xrightarrow[\mathcal L_{(k,k+2)}]{{\rm condense}}\mathbb C^{0|1}\,\mathcal L_{0}+\mathbb C^{0|1}\,\mathcal L_{2}\,.
\end{align}

\section{$\mathcal N=1$ minimal models}
\label{App:N=1}
\begin{figure}[H]
\centering
    \includegraphics[width=0.5\textwidth]{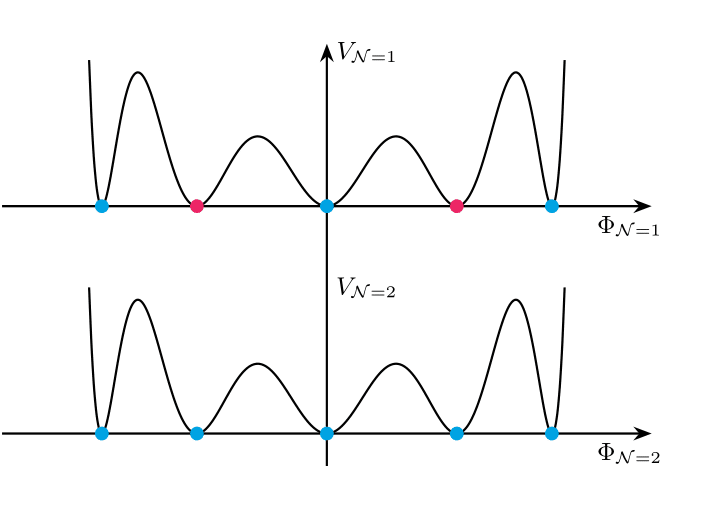}
\caption{$SM^{\rm def}_{6,8}$ and $A_4$, where the red dots denote fermionic vacua, and the blues for bosonic ones.}
\label{fig:N_1_N_2}
\end{figure}

One can show that the 't Hooft anomalies of the superfusion category exactly encode the delicate difference between the vacua types in the $\mathcal N=1,\, 2$ models. In fact, the analysis on $\mathcal N=2$ straightforwardly extends to other $\mathcal N=1$ minimal models with the least relevant deformation. The superfusion category of the deformed $\mathcal N=1$ models $SM_{k+2,k+4}^{\rm def}$ is captured by another variant of ${\rm SU(2)}_k$, $\mathscr C^{\,\mathcal N=1}_{A_{k+1}}=\{\,{\mathcal L}_a\,|a=0,1,2,\dots k\}$, where $\mathcal L_a$ is a $q$-type ($m$-type) object when $a$ is odd (even). It results in a dramatic change to the $\mathcal N=1$ vacua structure that the number of differences between the bosonic and fermionic vacua is at most 1, depending on $k$ even or odd, i.e.
\begin{equation}
    I_W^{\mathcal{N}=1}(A_{k+1}^{\textrm{def}})= N_{m\text{-type}}-N_{q\text{-type}}=
    \begin{cases}
    0\,, \quad {\rm for}\ \ k\ \ {\rm odd}\\
    1\,, \quad {\rm for}\ \ k\ \ {\rm even}\\
    \end{cases}
\end{equation}
More details of $\mathcal N=1$ models are presented in a companion paper \cite{toappear}. We summarize the difference of vacua structures between the $\mathcal N=1$ and $\mathcal N=2$ models in FIG.~\ref{fig:N_1_N_2}.

%
  %
 %
%
%
%
  %
 %
%
   %

\section{Mixed 't Hooft anomaly from superpentagon equations}\label{App:pentagon_anomaly}
In this appendix, we first briefly explain the $\mathbb Z_8^\nu$ classification of a $\mathbb Z_2$-symmetry in a fermionic system \cite{Chang:2022hud}. The simplest way to see it is to consider $\nu$ copies of Majorana fermions. The $\mathbb Z_2$-symmetry can be realized by the $\eta_\nu\equiv(-1)^{F_L}$, acting on the tensor products of $\nu$-copies of Majorana fermions. The $\mathbb Z_8$ classification of $\eta_\nu$ can be seen from the partition function,
\begin{align}
    \mathcal Z_{\nu}=|\chi_0+\chi_{1/2}|^{2\nu}\,,
\end{align}
where $\chi_0$, $\chi_{1/2}$, and $\chi_{1/16}$ are the standard characters for the vacuum, energy density, and spin primaries in Ising model. Inserting the $\eta_{\nu}$-line along the time direction, the defect partition function is spelled as
\begin{align}
    \mathcal Z^{\eta_{\nu}}_\nu=(\bar\chi_0+\bar\chi_{1/2})^\nu\chi_{1/16}^\nu\,.
\label{eq:defect_pf}
\end{align}
Therefore the defect Hilbert space $\mathcal H_{\eta_\nu}$ contains states of spin $\frac{\nu}{16}+\frac{1}{2}\mathbb Z$. Clearly, the spin selection rules admit a $\mathbb Z_8$-periodicity. We can classify the $\mathbb Z_2$ symmetry in terms of its spin selection rules, i.e., $\nu$. For $\nu=1,3,5,7$, the $\eta_{\nu}$ is a $q$-type object. On the other hand, for $\nu=0,2,4,6$, the $\eta_\nu$-line is of $m$-type. The $\nu=0$ and $\nu=4$ cases are corresponding to the ordinary non-anomalous and anomalous $\mathbb Z_2$-symmetry.

In our current case, we are interested in the following super-fusion category $\tilde{\mathscr C}_{A_2}=\mathbb Z_2^F\times \mathbb Z_2^\eta$. Here $\mathbb Z_2^F$ is the fermion parity symmetry $(-1)^F$, whereas $\mathbb Z_2^\eta$ corresponding to a chiral rotation for the $\mathcal N=2$ chiral multiplet $\Phi=(\phi,\,\psi)$,
\begin{equation}
        \eta:\quad\phi \rightarrow -\phi\,,\quad \psi\rightarrow \gamma_5 \psi\,,
\end{equation}
is spontaneously broken. Here, different from eq.~\eqref{eq:N=1_F_L}, $\phi$ is a complex boson and $\psi$ a Dirac fermion. Therefore all the symmetric lines $I$, $(-1)^F$, $\eta$, and $\eta(-1)^F$ are of $m$-type, but the fusions between $\eta$ and $(-1)^F\eta$ have non-trivial fermionic junctions summarized as follows:
\begin{align}
    &\eta\cdot\eta=\mathbb C^{0|1}I\,,\quad
    (-1)^F\eta\cdot (-1)^F\eta =\mathbb C^{0|1}I\,,\quad \notag\\
    &{\rm and}\quad
     (-1)^F\eta\cdot \eta =\eta\cdot  (-1)^F\eta= \mathbb C^{0|1}(-1)^F\,.
\end{align}
Using these data, one can determine the mixed 't Hooft anomaly between $(-1)^F$ and $\eta$, encoded in their $F$-moves,
 \begin{align}
     &\begin{gathered}
 \begin{tikzpicture}[scale=.75]
 \draw [line,dashed,ultra thick,MidnightBlue,->-=.55] (0,0)  -- (0,-1) node [below=-3pt] {$\cL_4$};
 \draw [line,dashed,ultra thick,BurntOrange,-<-=.55] (-1,1) -- (-2,2) node [above=-3pt] {$\cL_1$};
 \draw [line,dashed,ultra thick,ForestGreen,-<-=.55] (0,0) -- (-1,1) ;
 \draw [line,dashed,ultra thick,BurntOrange,-<-=.55] (0,0) --(2,2) node [above=-3pt] {$\cL_3$};
 \draw [line,dashed,ultra thick,MidnightBlue,-<-=.55] (-1,1) --(0,2) node [above=-3pt] {$\cL_2$};
 \draw (-.5,.5) node [ForestGreen,above right=-3pt] {$\cL_5$};
 \draw (0,-0.15) node [left]{\textcolor{black}{$\beta$}};
 \draw (-1,1) node [below]{\textcolor{black}{$\alpha$}};
 \end{tikzpicture}
 \end{gathered}\\~=~\sum_{\cL_6,\delta,\gamma} &\mathcal F_{\cL_4}^{\cL_1 \cL_2 \cL_3}[\cL_5,\cL_6]^{\alpha\beta}_{\delta\gamma}
 \begin{gathered}
 \begin{tikzpicture}[scale=.75]
 \draw [line,dashed,ultra thick,MidnightBlue,->-=.55] (0,0)  -- (0,-1) node [below=-4pt] {$\cL_4$};
 \draw [line,dashed,ultra thick,BurntOrange,-<-=.55] (0,0) -- (-2,2) node [above=-3pt] {$\cL_1$};
 \draw [line,dashed,ultra thick,ForestGreen,-<-=.55] (0,0) -- (1,1) ;
 \draw [line,dashed,ultra thick,BurntOrange,-<-=.55] (1,1) --(2,2) node [above=-3pt] {$\cL_3$};
 \draw [line,dashed,ultra thick,MidnightBlue,-<-=.55] (1,1) --(0,2) node [above=-3pt] {$\cL_2$};
 \draw (.5,.5) node [ForestGreen,above left=-3pt] {$\cL_6$};
 \draw (0,-0.15) node [right]{\textcolor{black}{$\delta$}};
 \draw (1,1) node [below]{\textcolor{black}{$\gamma$}};
 \end{tikzpicture}
 \end{gathered}\,,\notag
 \end{align}
 where we use a compact notation $\alpha=(\alpha^{\bf b},\alpha^{\bf f})$. Consistency yields the famous superpentagon equations,
\begin{align}
\label{eq:superpentagon}
    &\sum_\varepsilon \mathcal F^{icd}_e[j,k]^{\beta\chi}_{\delta\varepsilon}\, \mathcal F^{abk}_e[i,l]^{\alpha\varepsilon}_{\phi\gamma}\\
    &= (-1)^{s(\alpha) s(\delta)} \sum_{m\kappa\eta\iota} \mathcal F^{abc}_j[i,m]^{\alpha\beta}_{\eta\iota}\,\mathcal F^{amd}_e[j,l]^{\iota\chi}_{\kappa\gamma}\,\mathcal F^{bcd}_l[m,k]^{\eta\kappa}_{\delta\phi}\,,\notag  
\end{align}
where $\cL_i$ is abbreviated to $i$ (and similarly for other indices). It turns out that these equations, together with a choice of gauge, severely constrain the possible values of $\mathcal F$-symbols. The number of solutions to \eqref{eq:superpentagon} is finite, and each of them is thus rigid against local deformations.

One can find 8 solutions by solving \eqref{eq:superpentagon}. The 8 solutions, as explained in the beginning of the section, can be labeled by the spin selection rules of $(-1)^F$, $\eta$ and $(-1)^F\eta$, summarized in a triplet $(\nu_F,\nu_\eta,\nu_{F\eta})$, where $\nu_F=0,4$ and $\nu_\eta,\,\nu_{F\eta}=2,6$ \cite{Thorngren:2021yso, Chang:2022hud}. For our case, $(-1)^F$ need to be non-anomalous, and thus $\nu_F=0$. On the other hand, we need to choose $\nu_\eta=2$ as it acts on a Dirac fermion, or equivalently two Majorana fermions. The $\nu_{F\eta}$ can be therefore fixed by the following constraint \cite{note:CCX},
\begin{align}
    \nu_\eta+\nu_{F\eta}\equiv 0\,({\rm mod}\ 8)\,\quad\Longrightarrow\quad
    \nu_{F\eta}=6\,.
\end{align}
The above constraint is easily understood by studying the partition functions of two copies of Majorana fermions in presence of symmetric lines $\eta$ or $(-1)^F\eta$ along the temporal direction. One finds that
\begin{align}
     &\mathcal Z_{\nu=2}^\eta=(\bar\chi_0+\bar\chi_{1/2})^2\chi_{1/16}^2\,,\notag\\
     \quad{\rm and}\quad
     &\mathcal Z_{\nu=2}^{(-1)^F\eta}=\bar\chi_{1/16}^2(\chi_0+\chi_{1/2})^2\,.
\end{align}
Therefore we can read off the spin selection rules $\nu_\eta=2$ and $\nu_{F\eta}=6$. Although the spin selection rules are obtained in the context of CFTs, they are rigid against local deformations and can be applied to our gapped cases. 

Overall, the full $\mathcal F$-symbols in the superfusion category $\tilde{\mathscr C}_{A_2}$ are determined by the triplet $(\nu_F,\nu_\eta,\nu_{F\eta})=(0,2,6)$. In fact, for both two solutions $(\nu_F,\nu_\eta,\nu_{F\eta})=(0,2,6)$ or $(0,6,2)$, we have equation (\ref{eq:anomaly}), i.e. 
\begin{align}
    \mathcal F^{\,(-1)^F,\,\eta,\,(-1)^F}_{\eta}\left[(-1)^F\eta,\,(-1)^F\eta\right]^{00}_{00}=-1\,.
\end{align}

We anticipate that the above analysis can be straightforwardly generalized to the superfusion category $\tilde{\mathscr C}_{A_{k+1}}$ by solving more involved superpentagon equations. By specifying the spin-selection rule, one would be able to obtain the non-trivial mixed 't Hooft anomaly between $(-1)^F$ and $\mathcal L_1$.


As a final remark, if one wrongly assumes that the junction $V_{\eta,\eta,I}$ is bosonic, then the mixed 't Hooft anomaly between $\eta$ and $(-1)^F$ turns out to be trivial. One thus cannot find the fractional fermion number.

\section{Generalized Orbifoldings}\label{App: orbifold}
We explain more details on the $D/E$-type $\mathcal N=2$ minimal model $D_{k+2}$ with its least relevant deformation. First, the $D_{k+2}$ minimal model, obtained from its $A_{2k+1}$ cousin by gauging a $\mathbb Z_2$-symmetry: $\Phi\rightarrow -\Phi$, has superpotential
\begin{align}
    &\mathcal W_{A_{2k+1}}(\Phi)=\frac{1}{2k+2}\Phi^{2k+2}\notag\\
    \xrightarrow[\mathbb Z_2{\text{-gauging}}]{\Phi\rightarrow -\Phi,\,X\equiv\Phi^2}\,&\mathcal W_{D_{k+2}}(X,\,Y)=\frac{1}{2k+2}X^{k+1}+\frac{1}{2}XY^2\,.
\end{align}
Further notice that, in $A_{2k+1}$, the least relevant operator $\Phi_{(2k,2k)}=\Phi^{2k}+\cdots$ is invariant with respect to the $\mathbb Z_2$-symmetry. It is thus preserved under the $\mathbb Z_2$-orbifolding, and precisely the heaviest relevant operator $\sim X^k+\cdots$. We therefore conclude that the $D$-type least relevantly deformed model, $D_{k+2}^{\rm def}$, can be achieved via a $\mathbb Z_2$-gauging of $A_{2k+1}^{\rm def}$ all through along the RG-flow. On the categorical level, the to-be-gauged $\mathbb Z_2$-line in $A_{2k+1}^{\rm def}$ is ${\mathcal L}_{2k}\in \tilde{\mathscr C}_{A_{2k+1}}$. Together with $\mathcal L_0$, they form an algebraic object in $\tilde{\mathscr C}_{A_{2k+1}}$ known as Frobenius algebra,
\begin{align}
    \mathscr A_{D_{k+2}}\equiv\mathcal L_0\oplus\mathcal L_{2k}\,,
\end{align}
which defines a topological interface between the theories $A_{2k+1}^{\rm def}$ and $D_{k+2}^{\rm def}$
\begin{figure}[H]
    \centering
     \begin{tikzpicture}[scale=1.251]
        \draw [line,lightgray] (0,0)  -- (0,2) -- (2,2) -- (2,0) -- (0,0);
        \filldraw[MidnightBlue,opacity=0.2] (0,0)--(1,0)--(1,2)--(0,2)--(0,0);
        \filldraw[Cerulean,opacity=0.1] (1,0)--(2,0)--(2,2)--(1,2)--(1,0);
        \draw[ultra thick,Cerulean, ->-=.55] (2,0)--(2,2);
        \draw[ultra thick, double=white, draw=Teal] (1,0)--(1,2);
        \node at (.5, 1) {\footnotesize $D_{k+2}^{\rm def}$};
        \node at (1.5, 1) {\footnotesize $A_{2k+1}^{\rm def}$};
        \node at (2.4, -.3) {\footnotesize $\mathcal M_{\tilde{\mathscr C}_{A_{2k+1}}}$};
        \node at (1.1, -.3) {\footnotesize $\mathscr A_{D_{k+2}}$};
       \draw[sharp arrow, ultra thick, Teal] (2.5,1) -- (4,1) node[midway, above=2pt] {\footnotesize gauge $\mathscr A_{2k+1}$};
    \draw [line,lightgray] (4.3,0)  -- (4.3,2) -- (6.3,2) -- (6.3,0) -- (4.3,0);
    \filldraw [MidnightBlue,opacity=0.2] (4.3,0)  -- (4.3,2) -- (6.3,2) -- (6.3,0) -- (4.3,0);
        \draw[ultra thick, MidnightBlue, ->-=.55] (6.3,0)--(6.3,2);
        \node at (5.3, 1) {\footnotesize $D_{k+2}^{\rm def}$};
        \node at (6.3, -.3) {\footnotesize $\mathcal M_{\tilde{\mathscr C}_{D_{k+2}}}$};
    \end{tikzpicture}
    \caption{A boundary element in $\mathcal M_{\tilde{\mathscr C}_{D_{k+2}}}$ from the fusion of the interface $\mathscr A_{D_{k+2}}$ to a boundary element in $\mathcal M_{\tilde{\mathscr C}_{A_{2k+1}}}$} 
\end{figure}
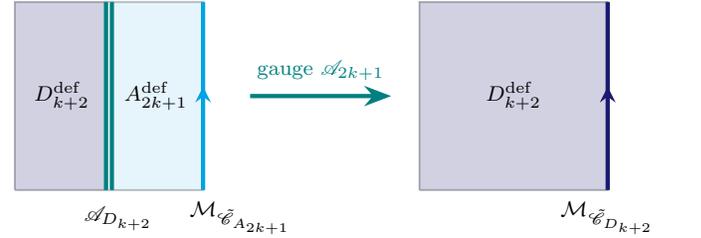
\noindent where $\tilde{\mathscr C}_{D_{k+2}}$ is the SSB superfusion category preserved in $D_{k+2}^{\rm def}$, and the associated modular category $\mathcal M_{\tilde{\mathscr C}_{D_{k+2}}}$ encodes the vacua for this theory. $\tilde{\mathscr C}_{D_{k+2}}$ can be computed via a ``bosonic condensation" of $\mathscr A_{D_{k+2}}$ in $\tilde{\mathscr C}_{A_{2k+1}}$,
\begin{align}
    D_{k+2}^{\rm def}=A_{2k+1}^{\rm def}/\mathbb Z_2\,,\quad
    \tilde{\mathscr C}_{D_{k+2}}=\tilde{\mathscr C}_{A_{2k+1}}\big/{\mathscr A}_{D_{k+2}}\,.
\end{align}
To demonstrate the condensation procedure, we compute the first non-trivial $\tilde{\mathscr C}_{D_4}$: Within $\tilde{\mathscr C}_{A_5}$ the action of ${\mathcal L}_4$ exchanges the lines in the pairs $({\mathcal L}_0,\, {\mathcal L}_4)$, $({\mathcal L}_1,\, {\mathcal L}_3)$, while fixing the line ${\mathcal L}_2$. Therefore, after condensing ${\mathcal L}_4$, ${\mathcal L}_2$ splits to two simple lines, denoted by $\widetilde{\mathcal L}_{\pm}$, and the two pairs condense to two simple lines $\widetilde{\mathcal L}_0\equiv ({\mathcal L}_0,\, {\mathcal L}_4)$ and $\widetilde{\mathcal L}_1\equiv ({\mathcal L}_1,\, {\mathcal L}_3)$ respectively. From the fusion rule \eqref{eq: su_2_N=2}, one can verify that lines $\widetilde{\mathcal L}_{0,\pm}$ furnish a $\mathbb Z_3$-symmetry. Together with $\widetilde{\mathcal L}_1$, one has
\begin{align}
    &\qquad\qquad \tilde{\mathscr C}_{D_4}=\left\{\widetilde{\mathcal L}_0,\, \widetilde{\mathcal L}_+,\, \widetilde{\mathcal L}_-,\, \widetilde{\mathcal L}_1\right\}\,, \quad {\rm with}\\
    &\widetilde{\mathcal L}_{0,\pm}\cdot \widetilde{\mathcal L}_1=\mathbb C^{1|0}\,\widetilde{\mathcal L}_1\,, \quad
   \widetilde{\mathcal L}_1\cdot \widetilde{\mathcal L}_1=\mathbb C^{0|1}\left(\widetilde{\mathcal L}_0+\widetilde{\mathcal L}_++\widetilde{\mathcal L}_-\right)\,,\notag
\end{align}
which can be regarded as a variant of $TY(\mathbb Z_3)$ with non-trivial fermionic junctions for $\widetilde{\mathcal L}_1\cdot \widetilde{\mathcal L}_1$.  

Since $\tilde{\mathscr C}_{D_4}$ is completely SSB, the $D$-type deformed model has Witten index $I_W(D_{4}^{\rm def})=4$. One repeats the superstrip algebra analysis, and finds that the only connected quiver representation corresponds to the element $\widetilde{\mathcal L}_1$:
\begin{figure}[H]
\centering
    \begin{tikzpicture}[
    dot_style/.style={circle, fill, inner sep=1.5pt},
    blue_arc/.style={Cerulean, thick, -Stealth, bend left=25},
    red_arc/.style={WildStrawberry, thick, -Stealth, bend left=25},
    label_below/.style={below=2pt}, scale=0.80063
]

\node at (-1, 0) {$\mathcal R_{\widetilde{\mathcal L}_1}$:};

    \node[dot_style, label={[label distance=5pt]below:$0$}] (n0) at (0, 0) {};
    \coordinate (n0pr) at (0.1, .125) {};
    \coordinate (n0pl) at (-0.9, .125) {};
    
    \coordinate (n0nr) at (0.1, -.125) {};
    \coordinate (n0nl) at (-0.9, -.125) {};

    \node[dot_style, label={[label distance=5pt]right:$1$}] (n1) at (1.5, 0) {};
    
    \coordinate (n1pl) at (1.4, .125) {};
    \coordinate (n1nl) at (1.4, -.125) {};
    
    \coordinate (n1pr) at (1.442, .149) {};
    \coordinate (n1nr) at (1.658, 0.024) {};

    \coordinate (n1pr2) at (1.658, -0.024) {};
    \coordinate (n1nr2) at (1.442, -.149) {};
    \node[dot_style, label={[label distance=3pt]right:$+$}] (n2) at (2.25, 1.299) {};
    \coordinate (n2pl) at (2.092, 1.275) {};

    \coordinate (n2nl) at (2.308, 1.150) {};
    \node[dot_style, label={[label distance=3pt]right:$-$}] (n3) at (2.25, -1.299) {};
    \coordinate (n3pl) at (2.308, -1.150) {};
        
    \coordinate (n3nl) at (2.092, -1.275) {};

    \draw[blue_arc] (n0pr) to (n1pl);
    \draw[red_arc] (n1nl) to (n0nr);
    \draw[red_arc] (n1pr) to (n2pl);
    \draw[blue_arc] (n2nl) to (n1nr);
    \draw[red_arc] (n1pr2) to (n3pl);
    \draw[blue_arc]  (n3nl) to (n1nr2);
    
\end{tikzpicture}
\end{figure}
\noindent The quiver representation encodes the soliton spectra of $D_{k+2}^{\rm def}$ as follows: There are three degenerate sets of fundamental (anti-)solitons connecting four vacua $|0\rangle$, $|\pm\rangle$ and $|1\rangle$ in a $D_4$-quiver pattern; each of them contains a pair of bosonic/fermionic one-particle states forming an irreducible $\mathcal N=2$ massive supermultiplet. 

The generalization to $D_{k+2}^{\rm def}=A_{2k+1}^{\rm def}/\mathbb Z_2$ is straightforward that the TDL ${\mathcal L}_k\in \tilde{\mathscr C}_{A_{2k+1}}$ splits to two $\widetilde{\mathcal L}_{\pm}\in \tilde{\mathscr C}_{D_{k+2}}$ which is completely SSB. There are thus $k+2$ vacua, among which the (anti-)soliton degeneracies are encoded in a $D_{k+2}$-quiver corresponding to the element $\widetilde{\mathcal L}_1\in \tilde{\mathscr C}_{D_{k+2}}$:
\begin{figure}[H]
\centering
    \begin{tikzpicture}[
    dot_style/.style={circle, fill, inner sep=1.5pt},
blue_arc/.style={Cerulean, thick, -Stealth, bend left=25}, red_arc/.style={WildStrawberry, thick, -Stealth, bend left=25}, label_below/.style={below=2pt}, scale=0.80063]

\node at (-1, 0) {$\mathcal R_{\widetilde{\mathcal L}_1}$:};

    \node[dot_style, label={[label distance=5pt]below:$0$}] (n0) at (0, 0) {};
    \coordinate (n0pr) at (0.1, .125) {};
    \coordinate (n0pl) at (-0.9, .125) {};
    
    \coordinate (n0nr) at (0.1, -.125) {};
    \coordinate (n0nl) at (-0.9, -.125) {};

    \node[dot_style, label={[label distance=5pt]below:$1$}] (n1) at (1.5, 0) {};
    \coordinate (n1pr) at (1.6, .125) {};
    \coordinate (n1pl) at (1.4, .125) {};
    
    \coordinate (n1nr) at (1.6, -.125) {};
    \coordinate (n1nl) at (1.4, -.125) {};
 
    \node[dot_style, label={[label distance=5pt]below:$2$}] (n2) at (3.0, 0) {};
    \coordinate (n2pr) at (3.1, .125) {};
    \coordinate (n2pl) at (2.9, .125) {};

    \coordinate (n2nr) at (3.1, -.125) {};
    \coordinate (n2nl) at (2.9, -.125) {};
  
    \node[dot_style, label={[label distance=5pt]below:$3$}] (n3) at (4.5, 0) {};
    \coordinate (n3pr) at (4.6, .125) {};
    \coordinate (n3pl) at (4.4, .125) {};
        
    \coordinate (n3nr) at (4.6, -.125) {};
    \coordinate (n3nl) at (4.4, -.125) {};

    \node at (5.25, 0) {$\dots$};

    \node[dot_style, label={[label distance=5pt]below:$k-2$}] (k2) at (6, 0) {};
    \coordinate (nk2pr) at (6.1, .125) {};
    \coordinate (nk2pl) at (-6.9, .125) {};
    
    \coordinate (nk2nr) at (6.1, -.125) {};
    \coordinate (nk2nl) at (-6.9, -.125) {};

\node[dot_style, label={[label distance=5pt]right:$k-1$}] (nk1) at (7.5, 0) {};
    
    \coordinate (nk1pl) at (7.4, .125) {};
    \coordinate (nk1nl) at (7.4, -.125) {};
    
    \coordinate (nk1pr) at (7.442, .149) {};
    \coordinate (nk1nr) at (7.658, 0.024) {};

    \coordinate (nk1pr2) at (7.658, -0.024) {};
    \coordinate (nk1nr2) at (7.442, -.149) {};

    \node[dot_style, label={[label distance=3pt]right:$+$}] (np) at (8.25, 1.299) {};
    \coordinate (nppl) at (8.092, 1.275) {};

    \coordinate (npnl) at (8.308, 1.150) {};

    \node[dot_style, label={[label distance=3pt]right:$-$}] (nn) at (8.25, -1.299) {};
    \coordinate (nnpl) at (8.308, -1.150) {};
        
    \coordinate (nnnl) at (8.092, -1.275) {};

    \draw[blue_arc] (n0pr) to (n1pl);
    \draw[red_arc] (n1nl) to (n0nr);
    \draw[red_arc] (n1pr) to (n2pl);
    \draw[blue_arc] (n2nl) to (n1nr);
    \draw[blue_arc] (n2pr) to (n3pl);
    \draw[red_arc] (n3nl) to (n2nr);
    \draw[blue_arc] (nk2pr) to (nk1pl);
    \draw[red_arc] (nk1nl) to (nk2nr);
    \draw[red_arc] (nk1pr) to (nppl);
    \draw[blue_arc] (npnl) to (nk1nr);
    \draw[red_arc] (nk1pr2) to (nnpl);
    \draw[blue_arc]  (nnnl) to (nk1nr2);
\end{tikzpicture}
\end{figure}
The same line of reasoning can be applied in parallel to study the least relevant deformation of the $E$-type $\mathcal N=2$ minimal model. First, the $E_{6,7,8}$ minimal CFTs can be achieved by gauging non-invertible symmetries in $A_{11, 17, 29}$, where there exist exceptional Frobenius algebra,
\begin{align}
    &\mathscr A_{E_6}=\mathcal L_0\oplus\mathcal L_6\,,\quad
    \mathscr A_{E_7}=\mathcal L_0\oplus\mathcal L_8\oplus\mathcal L_{16}\,,\notag\\
    &\mathscr A_{E_8}=\mathcal L_0\oplus\mathcal L_{10}\oplus\mathcal L_{18}\oplus\mathcal L_{28}\,,\quad
\end{align}
Since $\mathscr A_{E_{6,7,8}}$ are all preserved in the least relevantly deformed model $A_{11,17,29}^{\rm def}$, analogue to the analysis in $D$-type case, we are able to compute SSB superfusion category 
\begin{align}
\tilde{\mathscr C}_{E_{7,8,9}}=\tilde{\mathscr C}_{A_{11,17,29}}/\mathscr A_{E_{6,7,8}}\,,
\end{align}
preserved in $E_{6,7,8}^{\rm def}$, as well as the associated modular category $\mathcal M_{\tilde{\mathscr C}_{E_{7,8,9}}}$ encoding their vacua.

All $\tilde{\mathscr C}_{E_{n}}$ contain a simple object $\widetilde{\mathcal L}_1\equiv\mathcal L_1\otimes\mathscr A_{E_n}$, for $n=6,7,8$, which generates the whole superfusion categories. Other simple objects in $\tilde{\mathscr C}_{E_{7,8,9}}$ can be systematically determined by computing ${\rm Hom}_{\tilde{\mathscr C}_{A}}(\mathcal L_a,\,\mathcal L_a\otimes\mathscr A_{E})$ in a standard procedure \cite{Fuchs:2002cm}. 

Corresponding to $\widetilde{\mathcal L}_1$, we can find an $E$-type superstrip algebra representation, which encodes their vacua structure and degenerate (anti)-soliton spectra.
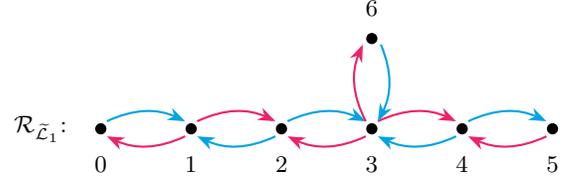
\begin{figure}[H]
\centering
    \begin{tikzpicture}[
    dot_style/.style={circle, fill, inner sep=1.5pt},
    blue_arc/.style={Cerulean, thick, -Stealth, bend left=25},
    red_arc/.style={WildStrawberry, thick, -Stealth, bend left=25},
    label_below/.style={below=2pt}, scale=0.80063
]

\node at (-1, 0) {$\mathcal R_{\widetilde{\mathcal L}_1}$:};

    \node[dot_style, label={[label distance=5pt]below:$0$}] (n0) at (0, 0) {};
    \coordinate (n0pr) at (0.1, .125) {};
    \coordinate (n0pl) at (-0.9, .125) {};
    
    \coordinate (n0nr) at (0.1, -.125) {};
    \coordinate (n0nl) at (-0.9, -.125) {};

    \node[dot_style, label={[label distance=5pt]below:$1$}] (n1) at (1.5, 0) {};
    \coordinate (n1pr) at (1.6, .125) {};
    \coordinate (n1pl) at (1.4, .125) {};
    
    \coordinate (n1nr) at (1.6, -.125) {};
    \coordinate (n1nl) at (1.4, -.125) {};

    \node[dot_style, label={[label distance=5pt]below:$2$}] (n2) at (3.0, 0) {};
    \coordinate (n2pr) at (3.1, .125) {};
    \coordinate (n2pl) at (2.9, .125) {};

    \coordinate (n2nr) at (3.1, -.125) {};
    \coordinate (n2nl) at (2.9, -.125) {};

    \node[dot_style, label={[label distance=5pt]below:$3$}] (n3) at (4.5, 0) {};
    \coordinate (n3pr) at (4.6, .125) {};
    \coordinate (n3pl) at (4.4, .125) {};
        
    \coordinate (n3nr) at (4.6, -.125) {};
    \coordinate (n3nl) at (4.4, -.125) {};

    \node[dot_style, label={[label distance=5pt]below:$4$}] (n4) at (6, 0) {};
    \coordinate (n4pr) at (6.1, .125) {};
    \coordinate (n4pl) at (5.9, .125) {};
        
    \coordinate (n4nr) at (6.1, -.125) {};
    \coordinate (n4nl) at (5.9, -.125) {};

    \node[dot_style, label={[label distance=5pt]below:$5$}] (n5) at (7.5, 0) {};
    \coordinate (n5pr) at (7.6, .125) {};
    \coordinate (n5pl) at (7.4, .125) {};
        
    \coordinate (n5nr) at (7.6, -.125) {};
    \coordinate (n5nl) at (7.4, -.125) {};

    \node[dot_style, label={[label distance=3pt]above:$6$}] (n6) at (4.5, 1.5) {};
    \coordinate (n6pl) at (4.35, 1.4) {};
        
    \coordinate (n6nl) at (4.65, 1.4) {};

    \draw[blue_arc] (n0pr) to (n1pl);
    \draw[red_arc] (n1nl) to (n0nr);
    \draw[red_arc] (n1pr) to (n2pl);
    \draw[blue_arc] (n2nl) to (n1nr);
    \draw[blue_arc] (n2pr) to (n3pl);
    \draw[red_arc] (n3nl) to (n2nr);
    \draw[red_arc] (n3pr) to (n4pl);
    \draw[blue_arc] (n4nl) to (n3nr);
    
    \draw[red_arc] (n3pl) to (n6pl);
    \draw[blue_arc] (n6nl) to (n3pr);

    \draw[blue_arc] (n4pr) to (n5pl);
    \draw[red_arc] (n5nl) to (n4nr);
\end{tikzpicture}
\caption{The $\mathcal R_{\widetilde{\mathcal L}_1}$ superstrip algebra representation for the least relevantly deformed $\mathcal N=2$ minimal model of $E_7$-type }
\end{figure}

\end{appendices}
\end{document}